\documentstyle[12pt]{article}
\def\fun#1#2{\lower3.6pt\vbox{\baselineskip0pt\lineskip.9pt
\ialign{$\mathsurround=0pt#1\hfil##\hfil$\crcr#2\crcr\sim\crcr}}}

\newcommand{\be}{\begin{equation}}
\newcommand{\ee}{\end{equation}}
\newcommand{\bd}{\begin{displaymath}}
\newcommand{\ed}{\end{displaymath}}
\newcommand{\ba}{\begin{array}}
\newcommand{\ea}{\end{array}}
\newcommand{\bt}{\begin{tabular}}
\newcommand{\et}{\end{tabular}}

\begin{document}
\vspace{1cm}
\begin{flushright}
UTPT-97-1
\end{flushright}
\begin{center}
{\Large\bf

THE LIGHT--FRONT MODEL FOR EXCLUSIVE\\[0.3cm]
SEMILEPTONIC $B$-- AND $D$--DECAYS}\\[1.3cm]

N. B. Demchuk, P. Yu. Kulikov, I. M. Narodetskii\\
Institute of Theoretical and Experimental Physics,\\
117259 Moscow, RF\\
and\\
Patrick J. O'Donnell,\\
Physics Department, University of Toronto,\\
Ontario M5S 1A7, Canada
\end{center}

\vspace{1cm}

\begin{abstract}
An explicit  relativistic  light-front  quark model is  presented
which gives the momentum transfer  dependent form factors of weak
hadronic  currents among heavy  pseudoscalar and vector mesons in
the whole accessible  kinematic region $ 0\leq q^2 \leq q^2_{max}
$.  For the numerical  investigations  of the $ B \to D^* l \nu_l
$, $ B \to  \rho l \nu_l $, $ D\to K^* l \nu_l $ and $ D \to \rho
l \nu_l $  semileptonic  decays  the  equal  time wave  functions
corresponding  to the  updated  version  of the  ISGW  model  are
adopted.  Using  the  available   experimental   information   on
branching fractions $ BR(B \to D^* l \nu_l) $ and $ BR(B \to \rho
l \nu_l)  $ the CKM  parameters  $ V_{cb}  $ and $  V_{ub} $ were
estimated:  $  |V_{cb}| = 0.036 \pm 0.004 $, $ |V_{ub}|  = 0.0033
\pm  0.0004 $.  The model is further  tested by  comparison  with
experimental data, QCD sum rules and lattice calculations.
\end{abstract}
\vspace{5cm}
\newpage

\section{Introduction}
The knowledge and  understanding  of the form factors of hadronic
currents is of decisive  importance for the  determination of the
quark mixing  parameters.  In  heavy--to--heavy  transitions  the
determination  of these form  factors  can be  obtained  based on
heavy-quark   symmetry   in  a  nearly   model--independent   way
\cite{NEU94}.  On the other hand, in the theoretical  description
of the heavy--to--light decays there is no symmetry one can apply
to constrain  the relevant  hadronic  form  factors.  Quark model
(QM)  calculations  can be very helpful for  heavy--to--heavy  as
well  as for  heavy--to--light  transitions  \cite{QM}.  Although
strict  theoretical  error limits cannot be given, they provide a
vivid  picture  of what is going on and  give  numerous  testable
predictions for quite different processes.

In refs.  \cite{DGNS96,GNS96}  the $ q^2 $ dependence of the form
factors among various  pseudoscalar mesons has been calculated in
the whole accessible kinematic region $ 0 \le q^2 \le q^2_{max} $
using  the  light-front  (LF)  QM.  The  calculations  have  been
performed  in a reference  frame where the  momentum  transfer is
purely  longitudinal.  The form factors were  determined by using
matrix  elements of the plus components of the currents which are
``good" operators in the LF formalism  \cite{TB}.  The purpose of
this   paper   is   to   generalize    the   results   of   refs.
\cite{DGNS96,GNS96}  and extend them for the  calculations of the
form factors governing  pseudoscalar--vector ($PS-V$) transitions
to $ 1^- $ states.  We will  show that the  latter  can  again be
expressed in terms of the matrix  elements of the plus components
of the currents in a reference frame  corresponding to the purely
longitudinal momentum transfers, but the calculations require the
use of a limiting procedure.

The  plan  of the  paper  is as  follows.  In  Sections  2--4  we
describe our approach for the  calculation of the weak meson form
factors.  In Section 5 the numerical results for the form factors
and  the  exclusive  decay  rates  of  the  $B  \to  D(D^*)  \ell
\nu_{\ell}$,  $B \to \pi(\rho)  \ell  \nu_{\ell}$,  $D \to K(K^*)
\ell  \nu_{\ell}$  and $D \to  \pi(\rho)  \ell  \nu_{\ell}$  weak
decays are presented and compared with  experimental data as well
as with the  results of other  approaches.  Section 6 contains  a
brief summary.

\section{Kinematics}
We denote by $ P_1 $, $ P_2 $ and $ M_1 $, $ M_2 $ the 4-momentum
and masses of the parent and daughter mesons,  respectively.  The
meson  states are denoted as $|P>$ for a  pseudoscalar  state and
$|P,\varepsilon>$  for a vector state, where $\varepsilon$ is the
polarization  vector,  satisfying   $\varepsilon\cdot  P=0$.  The
4-momentum  transfer  $q$  is  given  by $  q=P_1-P_2  $ and  the
momentum fraction $r$ is defined as

\be
r=\frac{P_2^+}{P_1^+}=1-\frac{q^+}{P^+_1}.
\ee

In the  following we work in the rest frame of the parent  meson.
The { \it  3}-momentum $ {\bf P_2} $ of the final meson is in the
plane  1-3, so that $ {\bf  q_{\bot}}  \not= 0 $.  We denote  the
angle  between $ {\bf P_2} $ and the 3-axis by $ \alpha $.  Then,
it can be easily verified that

\be
q^2=(1-r)(M_1^2-\frac{M_2^2}{r})-\frac{q^2_{\bot}}{r},
\ee

\be
q^2_{\bot }=M_2^2 (y^2-1) {\rm sin}^2\alpha,
\ee
where the``velocity  transfer" $ y $ is defined as $ y = u_1\cdot
u_2 $ with $ u_1 $ and $ u_2 $  being  the  4-velocities  of  the
initial and final mesons.  The relation between $ y $ and $ q^2 $
is given by

\be
 y =\frac{M^2_1+M^2_2-q^2}{2M_1M_2}.
\ee
The momentum  fraction of Eq.  (1) is invariant  under the boosts
along  the $ 3- $ axis and the  rotations  around  this  axis but
depends explicitly on the recoil direction.  Solving Eq.  (2) for
$r$ one obtains

\be
r(q^2,\alpha )=\zeta (y +\sqrt{y^2-1}{\rm cos} \alpha),
\ee
where $ \zeta=\frac{M_2}{M_1} $.  At the point of zero recoil $r$
does not depend on $\alpha $, $ r(q^2_{max})=\zeta $.

The hadronic form factors for semileptonic  decays are defined as
the   Lorentz--invariant   functions  arising  in  the  covariant
decomposition   of  matrix  elements  of  the  vector  and  axial
currents.  We define  the form  factors of the $ P_1(Q_1  \bar q)
\to  P_2(Q_2  \bar q) $  transitions  between  the  ground  state
$S$--wave    mesons   in   the   usual    way.   The    amplitude
$<P_2|V_{\mu}|P_1>  = <P_2|\bar  Q_2\gamma_{\mu}Q_1|P_1>$  can be
expressed in terms of two form factors

\be
<P_2|V_{\mu}|P_1>=\left(P_{\mu}-\frac{M^2_1-M^2_2}{q^2}
q_{\mu}\right)F_1(q^2)+\frac{M^2_1-M^2_2}{q^2}q_{\mu}F_0(q^2),
\ee
where $ P=P_1+P_2 $.  There is one form factor for the  amplitude
$ <P_2,\varepsilon|V_{\mu}|P_1> $

\be
<P_2,\varepsilon|V_{\mu}|P_1>=\frac{2i}{M_1+M_2}
\varepsilon_{\mu\nu\alpha\beta}\varepsilon^{*\nu}P^{\alpha}_1
P^{\beta}_2V(q^2),
\ee
and  three   independent   form  factors  for  the   amplitude  $
<P_2,\varepsilon|A_{\mu}|P_1>       =       <P_2,\varepsilon|\bar
Q_2\gamma_{\mu}\gamma_5Q_1|P_1> $

\begin{eqnarray}
<P_2,\varepsilon|A_{\mu}|P_1>& = &
\left((M_1+M_2)\varepsilon^{*\mu}
A_1(q^2)-\frac{\varepsilon^*q}{M_1+M_2}(P_1+P_2)_{\mu}A_2(q^2)-
2M_2\frac{\varepsilon^*q}{q^2}q_{\mu}A_3(q^2)\right) \nonumber \\
&&+2M_2
\frac{\varepsilon^*q}{q^2}q_{\mu}A_0(q^2),
\end{eqnarray}
where $ A_0(0) = A_3(0) $ and the form factor $A_3(q^2)$ is given
by the linear combination

\be
A_3(q^2)=\frac{M_1+M_2}{2M_2}A_1(q^2)-\frac{M_1-M_2}{2M_2}A_2(q^2).
\ee
Sometimes another set of form factors is introduced
\be
<P_2,\varepsilon|V_{\mu}|P_1> = ig(q^2)
\varepsilon_{\mu\nu\alpha\beta}\varepsilon^{*\nu}P^{\alpha}
q^{\beta},
\ee
\be
<P_2,\varepsilon|A_{\mu}|P_1> = -\left[f(q^2)\varepsilon^*_{\mu}
+a_+(q^2)(\varepsilon^*P)P_{\mu}+a_-(q^2)(\varepsilon^*P)q_{\mu}\right].
\ee
The relationship between two sets of form factors is given by

\be
V(q^2)=-(M_1+M_2)g(q^2),
\ee

\be
A_0(q^2)=-\frac{1}{2M_2}[f(q^2)+(M_1^2-M_2^2)a_+(q^2)+q^2a_-(q^2)],
\ee

\be
A_1(q^2)=-\frac{f(q^2)}{M_1+M_2},
\ee

\be
A_2(q^2)=(M_1+M_2)a_+(q^2).
\ee

In the massless  lepton limit the form factors  $F_0(q^2)$  and $
a_-(q^2)  $ do not  contribute  to the  decay  rates  because  of
leptonic current  conservation.  The differential decay width for
a semileptonic  decay of a $PS$ particle to another $PS$ particle
and a massless lepton pair ($e \bar\nu_e$ or $\mu\bar \nu_{\mu}$)
is given by

\be
\frac{d\Gamma}{dy}=\frac{G^2_FM^5_1}{12\pi^3}\zeta^4|V_{Q_1Q_2}|^2
(y^2-1)^{3/2}|F_1(y)|^2,
\ee
where $ V_{Q_1Q_2} $ is the CKM matrix element.  The differential
transition rate to a vector particle is

\be
\frac{d\Gamma}{dy}=\frac{G^2_FM_1}{48\pi^3}\zeta^2|V_{Q_1Q_2}|^2
(y^2-1)^{1/2}q^2(H^2_++H^2_-+H^2_0),
\ee
where the helicity amplitudes are given by

\be
H_0=\frac{M^2_1}{\sqrt{q^2}}\left((y-\zeta)(1+\zeta)A_1(y)-
2(y^2-1)\zeta\frac{A_2(y)}{1+\zeta}\right),
\ee

\be
H_{\pm}=M_1\left((1+\zeta)A_1(y)\mp2(y^2-1)^{1/2}\zeta
\frac{V(y)}{1+\zeta}\right).
\ee
Near the  kinematic  point  of zero  recoil,  $ y = 1$, Eq.  (17)
simplifies to \cite{DF93}

\be
\frac{d\Gamma}{dy}=\frac{G^2_FM_1^5}{16\pi^3}\zeta^2|V_{Q_1Q_2}|^2
(1-\zeta)^2(y^2-1)^{1/2}|A_1(q^2_{max})|^2,
\ee
while at  opposite  end of the heavy  meson  spectra,  namely, at
$q^2=0$ one obtains \cite{DT96}

\be
\frac{d\Gamma}{dy}=\frac{G^2_FM_1^5}{96\pi^3}|V_{Q_1Q_2}|^2\zeta
(1-\zeta^2)^3|A_0(0)|^2.
\ee

In the case of  heavy--to--heavy  transitions,  in the  limit  in
which the active quarks have infinite  mass, all the form factors
are given in terms of a single function $\xi(y)$, the Isgur--Wise
form factor.  In the realistic  case of finite quark masses these
relations are modified:  each form factor  depends  separately on
the dynamics of the process.

\section{The matrix elements of the vector and axial currents}

We determine all of the form factors by taking matrix elements of
the plus  components of the weak currents.  Applying the standard
LF    technique    for   the    time-like    momentum    transfer
\cite{DGNS96,GNS96} to the matrix element of the vector current $
\bar   Q_2\gamma^+Q_1  $  between  two  pseudoscalar  mesons  one
obtains\footnote{The  spectator quark carries the fraction $x$ of
the plus component of the meson  momentum,  while the heavy quark
carries the fraction $1-x$.}
\be
\ba{l}
J_V(q^2,r)=<P_2|\bar Q_2\gamma^+Q_1|P_1>=\int\limits^{r}_0
\frac{dx}{x}\int d^2 k_{\bot}\chi_2(x', k'^2_{\bot})
\chi_1(x, k^2_{\bot})
\cdot I_V,
\ea
\ee
where $I_V$ is the  contribution  of the Dirac  current and quark
spin structure:
\be
I_V=\mu_1\mu_2 Tr[R^+_{00}\bar u(\bar p_2,\lambda_2)
\gamma^+u(p_1,\lambda_1)R_{00}],
\ee
with
\be
\mu_1 = [M^2_{10}-(m_1-m)^2]^{1/2},
\ee
\be
\mu_2= [M^2_{20}-(m_2-m)^2]^{1/2},
\ee
and
 \be
    M_{10}^2  \equiv  M_{10}^2(x,  k^2_{\bot}) = \frac{m^2 +
     k^2_{\bot}}{x} + \frac{m^2_1 +  k^2_{\bot}}{1-x},
 \ee
 \be
    M_{20}^2  \equiv  M_{20}^2(x',  k'^2_{\bot}) = \frac{m^2 +
     k'^2_{\bot}}{x'} + \frac{m_2^2 +
 k'^2_{\bot}}{1-x'}. \\ \label{19}
 \ee
 In these  equations  $m_1$ and  $m_2$  are the  masses of  active
quarks, $m$ is the mass of the quark--spectator, $x'=\frac{x}{r}$
and ${\bf  k'_{\bot}}={\bf  k_{\bot}}+x'{\bf  q_{\bot}}$.  In our
kinematics
\be
k_1'= k_1-x'q_{\bot}, ~~~k_2'=k_2.
\ee
The  spin  wave  functions  $R_{JJ_3}$  can  be  found  in  refs.
\cite{Jold,DXT95}.  We choose a  phenomenological  wave  function
$\chi_i(x,  k^2_{\bot})$  by starting  with the  equal-time  wave
function $w_i(k^2)$ normalized according to

 \be
    \int_0^{\infty} dk ~ k^2 ~ w_i^2(k^2) = 1,
\ee
provided  that the fraction  $x$ is replaced by the  longitudinal
momentum $ k^{(1)}_3 $ defined as
 \be
    k^{(1)}_3 = \left( x - {1 \over 2} \right) M_{10} + \frac{m_1^2 -
    m^2}{2M_{10}},
 \ee
and the fraction $x'$ is replaced by the longitudinal  momentum $
k^{(2)}_3 $
 \be
    k^{(2)}_3 = \left( x' - {1 \over 2} \right) M_{20} + \frac{m_2^2 -
    m^2}{2M_{20}}.
 \ee
Explicitly, one has \cite{DGNS96}
 \be
    \chi_i(x,  k^2_{\bot}) = \frac{1}{2 (1 - x)} \frac{\sqrt{M_{i0} [1 -
    (m_i^2 - m^2)^2 / M^4_{i0}]}} {\sqrt{M^2_{i0} - (m_i - m)^2}} ~
    \frac{w_i(k^2)}{\sqrt{4\pi}}
 \ee
with $k^2 \equiv  k_{\bot}^2 + k_3^2$.  This procedure amounts to
a series  of  reasonable  (but  naive)  guesses  about  what  the
solution   of   a   relativistic   theory   involving   confining
interactions might look like.

The matrix element of the vector current $ \bar  Q_2\gamma^+Q_1 $
between pseudoscalar and vector mesons is given by 
\be
J^{(+1)}_V(q^2,q_{\bot})=
<P_2\varepsilon(+1)|\bar Q_2\gamma^+Q_1|P_1>=\int\limits^{r(q^2,q_{\bot})}_0
\frac{dx}{x}\int d^2k_{\bot}\chi_2(x', k'^2_{\bot})
\chi_1(x, k^2_{\bot})
\cdot I_V^{(+1)},
\ee
where

\be
I_V^{(+1)}=\mu_1 \mu_2 [Tr{R^+_{1+1}\bar u(\bar p_2,\lambda_2)\gamma^+u(p_1,
\lambda_1)R_{00}}]
\ee
To define the axial form  factors we need two matrix  elements of
the axial current $ \bar  Q_2\gamma^+\gamma_5Q_1  $ corresponding
to  the  transverse  and  longitudinal   polarization   states  $
\varepsilon(+1) $ and $ \varepsilon(0) $ of a vector meson
\be
J^{(+1)}_A(q^2,q_{\bot})=<P_2\varepsilon(+1)|\bar
Q_2\gamma^+\gamma_5Q_1|P_1>=\int\limits^{r(q^2,q_{\bot})}_0
\frac{dx}{x}\int d^2k_{\bot}\chi_2(x', k'^2_{\bot})
\chi_1(x, k^2_{\bot})
\cdot {\it I}_A^{(+1)},
\ee
\be
J^{(0)}_A(q^2,q_{\bot})=
<P_2\varepsilon(0)|\bar Q_2\gamma^+\gamma_5Q_1|P_1>=
\int\limits^{r(q^2,q_{\bot})}_0
\frac{dx}{x}\int d^2k_{\bot}\chi_2(x', k'^2_{\bot})
\chi_1(x, k^2_{\bot})
\cdot {\it I}_A^{(0)},
\ee
where

\be
I_A^{(+1)}=\mu_1 \mu_2Tr[R^+_{1+1}
\bar u(\bar p_2,\lambda_2)\gamma^+\gamma_5u(p_1,\lambda_1)R_{00}],
\ee

\be
 I_A^{(0)}=\mu_1 \mu_2 Tr[R^+_{10}
\bar u(\bar p_2,\lambda_2)\gamma^+\gamma_5u(p_1,\lambda_1)R_{00}].
\ee

Noting  that  $\bar   u(p_2,\lambda_2)\gamma^+   u(p_1,\lambda_1)
=\sqrt{4p_1^+p_2^+}\delta_{\lambda_2      \lambda_1}$,      $\bar
u(p_2,\lambda_2)\gamma^+\gamma_5                 u(p_1,\lambda_1)
=\sqrt{4p_1^+p_2^+}\varphi^+_{\lambda_2}                 \sigma_3
\varphi_{\lambda_1}$,  where $ \varphi_{\lambda}  $ are the Pauli
spinors and using the explicit  expressions  for light-cone  spin
structure functions $R_{JJ_3}$ from \cite{Jold,DXT95} we obtain
\be
I_V(x;r;k_{\bot};q_{\bot})=\frac{2M_1}{x'}[{\cal A}_1{\cal A}_2+{\bf k_{\bot}k'_{\bot}}],
\ee
\be
I^{(+1)}_V(x;r;k_{\bot};q_{\bot})=-\frac{\sqrt2M_1}{x'}
[k'_1{\cal A}_1-k_1{\cal A}_2+\frac{2k^2_2(k'_1-k_1)}{M_{20}+m+m_2}],
\ee
\be
I_A^{(+1)}(x;r;k_{\bot};q_{\bot})=
\frac{\sqrt2M_1}{x'}[(2x'-1)k'_1{\cal A}_1+k_1{\cal A}_2+2k'_1
\frac{{\cal A}_1{\cal B}+{\bf k_{\bot} k'_{\bot}}}{M_{20}+m+m_2}],
\ee
\be
I_A^{(0)}(x;r;k_{\bot};q_{\bot})=\frac{2M_1}{x'}[2x'(1-x')M_{20}{\cal A}_1+
\frac{(1-2x')M_{20}+m_2-m}{M_{20}+m+m_2}
({\bf k_{\bot}k'_{\bot}}+{\cal A}_1{\cal B})],
\ee
with

\be
 {\cal A}_1=xm_1+(1-x)m, ~~ {\cal A}_2=x'm_2+(1-x')m,
\ee
and
\be
 {\cal B}=(1-x')m-x'm_2.
\ee

\section{The definition of the vector and axial form factors}

We  first   reexamine  an  approach  of  ref.  \cite{DGNS96}   to
calculate  the  form  factors  $  F_1(q^2)  $ for  the $ PS-PS  $
transitions.  Eq.  (6)  for  the  plus  component  of the  vector
current  yields only one  constraint  for the two form  factors $
F_1(q^2)  $ and $  F_0(q^2)  $.  In order to invert  Eq.  (6) the
matrix element of the current was calculated in \cite{DGNS96}  in
two reference frames having the 3-axis parallel and anti-parallel
to the {\it 3}-momentum of the daughter meson.  This  corresponds
to the choice  $\alpha_1=0,\alpha_2=\pi$ where the angle $ \alpha
$ is  defined in Section 2.  One might ask what  happens  when we
use  two  other  frames.  Specifying  these  frames  by  the  two
arbitrary  angles $  \alpha_1 $ and $ \alpha_2 $ we can now write
Eq.  (35) of ref.  \cite{DGNS96} as

\be
F_1(q^2)=\frac{(1-r(\alpha_2))J_V(q^2,r(\alpha_1))-
(1-r(\alpha_1))J_V(q^2,r(\alpha_2))}
{2M_1(r(\alpha_1)-r(\alpha_2))}.
\ee
In   this   equation   we  can   take   the   limit  $   \alpha_2
\to\alpha_1=\alpha  $ which would  correspond  to the use of only
one reference  frame  specified by the angle $ \alpha $.  In this
case the form  factor $ F_1(q^2) $ is  expressed  in terms of the
matrix  element of the current and its  derivative.  Indeed, from
Eq.  (45) one derives 

\be
F_1(q^2)=\frac{1}{2M_1}\left[J_V(q^2,r(\alpha))+
(1-r(\alpha))\frac{\partial J_V(q^2,r(\alpha))}{\partial r}\right].
\ee
It is understood  that when  calculating  the partial  derivative
$\frac{\partial  J_V}{\partial  r} $  the  variable  $q_{\bot}  $
entering  the  integrand of Eq.  (22) is  expressed in terms of $
q^2 $ and $ r $ using Eq.  (2).

The  final  choice  of $ \alpha $ can be made by the  requirement
that at the  point of the  maximum  recoil,  $ q^2=0 $, Eq.  (46)
must  coincide  with the  standard  LF result  for $ q^2 \leq 0 $
obtained  in the Breit  frame  \cite{Jold,DXT95}.  One can easily
verify that this  condition  is fulfilled  only for $ \alpha=0 $.
Eq.  (46)  for  $  \alpha=0  $  coincides   with  that  of  refs.
\cite{DGNS96,GNS96}  at $q^2=0$ and $q^2=q^2_{max}$,  but the two
definitions differ at intermediate  $q^2$.  Recall  \cite{DGNS96}
that the  time-like  LF result for the ``good"  component  of the
weak  vector   current  $  J^+=J^0+J^3   $  coincides   with  the
contribution  of  the  spectator  pole  of the  Feynman  triangle
diagram, which  corresponds  to the valence quark  approximation,
while the remaining part of the Feynman  diagram, the so-called Z
graph, can not be expressed  directly in terms of a valence quark
wave  function.\footnote  { Note  that  the  contribution  of the
Z-graph is expressed in terms of the integral over the region $ r
\leq x \leq 1 $, therefore to diminish its  contribution one must
choose $ r $ (at given $ q^2 $) to have its maximum  close to $ 1
$; in this way we again  come to the choice $ \alpha = 0 $.}  The
sum of both  contributions  does not  depend,  of  course, on the
choice of the  frame but each  contribution  is  frame-dependent.
Therefore the time-like LF result for the form factors  generally
depends on the recoil direction of the daughter meson relative to
the 3-axis e.g.  on the choice of angles $ \alpha $ specifying  a
reference  frame.  Fortunately  this  dependence is marginal even
for  heavy-to-light  transitions.  We shall illustrate this point
later on.

Now we apply the same  procedure to the  calculation  of the form
factors  governing  the $ PS-V $  transitions.  The  vector  form
factor $ V(q^2) $ can be found from Eqs.  (7, 33).  Note that for
the pure  longitudinal  configuration  both the  integral  of Eq.
(33) and the covariant
\be
 \varepsilon_{+\nu \alpha \beta}\varepsilon^{\ast \nu}(+1)P^{\alpha}_1
P^{\beta}_2=
-\frac{i}{\sqrt2} M_1M_2\sqrt{y^2-1}{\rm sin} \alpha \equiv -\frac{i}{\sqrt2}
 M_1q_{\bot}
\ee
vanish.  Therefore  to  calculate  the form  factor $ V(q^2) $ we
will choose $ \alpha \not= 0 $ at the outset and let $ \alpha \to
0 $ at the end of calculations to get
\be
V(q^2)=\frac{1}{\sqrt2}(1+\zeta)\left[ \frac{\partial}{\partial q_{\bot}}
J^{(+1)}_V(q^2,q_{\bot})\right]_{q_{\bot}=0}.
\ee
To define the axial form factors we first note that
\be
J^{(+1)}_A=\frac{1}{\sqrt2}
M_1q_{\bot}(\frac{1+r(\alpha)}{r(\alpha)}a_+(q^2)+
\frac{1+r(\alpha)}{r(\alpha)}a_-(q^2)).
\ee
Then we proceed  exactly in the same way as was  described  above
for  calculating  the form  factor $ F_1(q^2)  $.  An  additional
difficulty is that the integral of Eq.  (35) also  vanishes for $
\alpha=0 $.  Therefore  to invert  Eq.(35) we first  solve it for
$\alpha=\alpha_1$ and $\alpha=\alpha_2$ and then take the limit $
\alpha_1,\alpha_2 \to 0 $.  We obtain
\be
a_{\pm}(q^2)=\pm \frac{1}{\sqrt2M_1}
({\it A^{(+)}}(r)+r(1\mp r)\frac{d{\it A^{(+)}}(r)}{dr}),
\ee
where
\be
 {\it A^{(+)}}(r)=
\left[ \frac{\partial}{\partial q_{\bot}}
J^{(+1)}_A(q^2,q_{\bot})
\right]_{q_{\bot}=0}.
\ee
In order to calculate the remaining  form factor  $f(q^2)$ we use
the longitudinal polarization $ \varepsilon(0)
 $ \cite{Jold} to get

\be
f(q^2)=-\frac{\zeta}{r}
\left[J^{(0)}_A(q^2,q_{\bot})+
\frac{M_2}{\sqrt2\zeta^2}
(r^2-\zeta^2)A^{(+)}(r)\right]_{q_{\bot}=0}.
\ee

At the point of the maximum  recoil one obtains  from Eqs.  (48), (50)
and( 52)
\be
V(0)=\frac{1}{\sqrt2}(1+\zeta)\left[\frac{d}{dq_{\bot}}
J^{(+1)}_V(0,q_{\bot})\right]_{q_{\bot}=0},
\ee
\be
A_0(0)=\frac{1}{2M_1}J^{(0)}_A(0,0),
\ee
and
\be
A_2(0)= \frac{1+\zeta}{\sqrt2}
\left[\frac{d}{dq_{\bot}}
J^{(+1)}_A(0,q_{\bot})
\right]_{q_{\bot}=0},
\ee
where the  relationships  (13, 15) between the form factors $ A_0
$, $ A_2 $ and the form  factors $ a_{\pm}  $ and $ f $ have been
used.  We show in the Appendix that Eqs.  (53-55)  coincide  with
the  standard   result   \cite{Jold,DXT95}   obtained  using  the
Hamiltonian  LF  formalism  in the Breit  frame\footnote{We  have
compared our numerical  results for the $ PS \to V $ form factors
with  those  of Jaus  \cite{Jnew}  using  the set of quark  model
parameters (the constituent quark masses and harmonic  oscillator
lengths)  found in  \cite{Jnew} by exploiting the duality of the
VMD and CQM pictures.  This  comparison is presented in Table 1.
Note that  there is a  misprint  in Table VI of ref. \cite{Jnew}.
The correct Table is obtained if the symbols $ V $, $ A_1 $ and $
A_2 $ are  replaced by $ A_1 $, $ A_2 $, and $ V $ .  Our results
for the vector and axial form  factors for $B$ and $D$ decays are
very  similar to the  results of ref.  \cite{Jnew}  but the rates
are a bit  different  and  reflect  the  difference  in the $q^2$
dependence of the form factors.}.

\section{Results}

The  calculation  of the  semileptonic  form factors  $F_1(q^2)$,
$V(q^2)$,   $A_0(q^2)$,   $A_1(q^2)$  and  $A_2(q^2)$   has  been
performed  using  our LF  results  (Eqs.  46,  48,  50,  52)  for
computing  the  matrix  elements  of the  vector  and axial  weak
currents.  The LF meson wave functions  $\chi_i$ are  constructed
according  to the rules given in Section 3.  For the radial  wave
functions appearing in Eq.  (32), the Gaussian ansatz of the ISGW
model  has been  adopted  in  which  the main  characteristic  is
confinement\footnote{There  is some model dependence  here; other
choices  \cite{GNS96}  for $  w(k^2)  $ will  lead  for  somewhat
different  predictions.}.  The values of the parameters are taken
from   the   updated    version    \cite{IS95}    of   the   ISGW
model\footnote{In  what  follows  this  updated  version  will be
referred as the ISGW2  model.}.  We use the  physical  values for
the meson  masses  taken  from the  latest  Particle  Data  Group
publication \cite{PDG96}.

Our results  for the form  factors at $ q^2=0 $ and  semileptonic
decay rates  $\Gamma$  (Eqs.  (16, 17)) are collected in Tables 1
and 2.  The $ q^2 $ behavior of the form factors  $F_1(q^2)$  for
$B \to D \ell  \nu_{\ell}$, $B \to \pi \ell \nu_{\ell}$, $D \to K
\ell  \nu_{\ell}$, and $D \to \pi \ell  \nu_{\ell}$  semileptonic
transitions  is shown in Figs.  1--4,  respectively,  while the $
q^2 $ behavior  of the form  factors $ V(q^2) $, $ A_0(q^2)  $, $
A_1(q^2) $, and $ A_2(q^2) $ for $B \to D^* \ell  \nu_{\ell}$, $B
\to \rho \ell  \nu_{\ell}$, $D \to K^* \ell  \nu_{\ell}$,  and $D
\to \rho \ell  \nu_{\ell}$  transitions  is shown in Figs.  5--8.
Note that near the zero-recoil  point $ y=1 $ the form factors do
not obey at all the pattern of pole dominance and even may have a
negative slope.  This behavior, especially  prominent for the $ b
\to u $  transitions,  can  be  ascribed  to the  fact  that  our
calculations  do not include the  contribution  of the quark-pair
creation  diagram  (the  Z-graph)  which,  in  another  language,
corresponds  to the  contribution  of  $|B^*\pi>$  or $ |D^*\pi>$
states  \cite{IW90}.  However, the corresponding  decay rates are
not affected by the Z-graph because the contribution to the rates
arising from the region near $ y=1 $ is kinematically suppressed.

In order to estimate  uncertainties  introduced  by a  particular
choice of the  reference  frame we have  compared  in Table 2 the
decay rates for the $ PS-PS $ transitions calculated from our new
Eq.  (46) (corresponding to the choice $ \alpha_1, \alpha_2 \to 0
$) and  from Eq.  (35) of ref.  \cite{DGNS96}  (corresponding  to
the choice $ \alpha_1 = 0,  \alpha_2 = \pi $).  We have found the
new  decay  rates  to be  slightly  smaller  than  those  of ref.
\cite{DGNS96}:  the differences comprise $ 7\% $ for $ b \to c $,
$ 9\% $ for $ b \to u $, $ 13\% $ for $ c \to s$, and $ 4\% $ for
$ c \to u $  transitions.  We conclude  that a good  stability of
the  results  is always  maintained  and the  decay  rates  (with
exception of $ c\to s $ case) are predicted with accuracy  better
than  $ 10\%  $.  The  same  conclusion  is  preserved  for  more
complicated  meson wave functions  corresponding to the inclusion
of the one-gluon exchange at small distances \cite{Gr96}.

We now discuss the results arranged in order of decreasing active
quark mass.  We will compare our results to experimental data and
to predictions of different models.

\subsection{Decays $B \to D \ell \nu_{\ell}$ and $B \to D^* \ell \nu_{\ell}$}

\indent  First, we want to  estimate  the  values of $ |V_{cb}| $
based on our predictions of the normalization of the form factors
and    their   $   q^2   $    dependence.   From   the    partial
widths\footnote{These  values have been derived by combining  the
branching  ratios  $Br(B^0 \to D^- \ell^+  \nu_{\ell}) = (1.9 \pm
0.5) \%$ and $Br(B^0 \to D^{-*}  \ell^+  \nu_{\ell})  = (4.56 \pm
0.27)  \%$  with  the  world   average  of  the  $B^0$   lifetime
$\tau_{B^0} = 1.56 \pm 0.06 ~ ps$  \cite{PDG96}.  Our  estimation
of $ \Gamma (B^0 \to D^{-*} \ell^+  \nu_{\ell}) $ agrees with the
recent result by CLEO collaboration \cite{Barish95} $ \Gamma (B^0
\to D^{-*}  \ell^+  \nu_{\ell}) = [2.99 \pm  0.19(stat)  \pm 0.27
(syst) \pm 0.20 (lifetime)] \cdot 10^{10} ~ s^{-1} $ based on the
average  of  the  LEP  and  CDF  measurements  of  the  lifetimes
\cite{VENUS94},  $\tau_{B^0} = 1.53 \pm 0.09 ~ ps$, $\tau_{B^+} =
1.68 \pm 0.12 ~ ps$.}
 \be
    \Gamma (B^0 \to D^- \ell^+ \nu_{\ell}) = (1.22 \pm 0.32) \cdot
    10^{10} ~ s^{-1},
 \ee
\be
    \Gamma (B^0 \to D^{-*} \ell^+ \nu_{\ell}) = (2.92 \pm 0.17) \cdot
    10^{10} ~ s^{-1}
 \ee
and our predicted rates for $\Gamma (B^0 \to D^- \ell^+ \nu_{\ell})$  and
$\Gamma (B^0 \to D^{*-} \ell^+ \nu_{\ell})$  from Table 2, we obtain
 \be
     |V_{bc}| = 0.037 \pm 0.004,~~(B^0 \to D^- \ell^+ \nu_{\ell}),
\ee
\be
     |V_{bc}| = 0.036 \pm 0.002,~~(B^0 \to D^{-*} \ell^+ \nu_{\ell}).
\ee

The  results are  consistent  in that the  extracted  values of $
V_{bc} $ are  consistent  for two decay modes.  Fig.9 shows the $
\frac{d\Gamma}{dq^2}    $   distribution    calculated    for   $
|V_{bc}|=0.036  $ and $  |V_{bc}|=0.0374  $.  The last  value has
been   obtained   from   our  fit  to  the   recent   CLEO   data
\cite{Barish95}.  Both LF predictions  are in agreement  with the
updated    ``experimental"     determinations    of    $|V_{bc}|$
\cite{Skwar95} obtained from exclusive and inclusive semileptonic
decays of the B-meson ($|V_{bc}|_{excl} = 0.0373 \pm 0.0045_{exp}
\pm 0.0065_{th}$ and  $|V_{bc}|_{incl}  = 0.0398 \pm 0.0008_{exp}
\pm  0.0040_{th}$), and also agree within the error bars with the
recent determination $|V_{bc}| = 0.041 \pm 0.003 \pm 0.002 $ from
a partial rate  $\frac{d\Gamma (B \to D^* \ell  \nu_{\ell})}{dy}$
corresponding  to the  kinematical  region  where $ D^* $  recoil
slowly  \cite{Richman96}.  In Table 3 we  compare  the  values of
$|V_{bc}|$ for several quark models.

Our   predicted   values   $   R_1=V(0)/A_1(0)=1.08   $   and   $
R_2=A_2(0)/A_1(0)=0.89   $  are  in  good   agreement   with  the
preliminary fit results by CLEO  \cite{Gronberg95} $ R_1=1.18 \pm
0.30  \pm 0.12 $ and $ R_2 = 0.71 \pm  0.22 \pm  0.07 $.  For the
branching ratios $ BR(B^0 \to D^- \ell^+  \nu_{\ell}) $, $ BR(B^0
\to D^{*-} \ell^+ \nu_{\ell}) $ we obtain

\be
 BR(B^0 \to D^- \ell^+ \nu_{\ell}) =2.12
\left|\frac{V_{cb}}{0.039} \right|^2,
\ee
\be
 BR(B^0 \to D^{*-} \ell^+ \nu_{\ell}) =5.40
\left|\frac{V_{cb}}{0.039} \right|^2.
\ee
These values only  slightly  exceed the  corresponding  branching
fractions obtained using the heavy-quark  symmetry  approximation
i.e.  adopting  a single  Isgur-Wise  function  for all the  form
factors:  $   BR(B^0   \to   D^-   \ell^+   \nu_{\ell})   =  1.65
\left|\frac{V_{cb}}{0.039}  \right|^2  $, and $ BR(B^0 \to D^{*-}
\ell^+ \nu_{\ell})  =5.17\left|\frac{V_{cb}}{0.039}  \right|^2 $.
Our  prediction  for  the  ratio  of  of  the  longitudinal   and
transversal  contributions to the $B \to D^*$ decay is 1.17 which
agrees   with   the   recent   result   by   CLEO   Collaboration
$\frac{\Gamma_L}{\Gamma_T}  = 1.24  \pm 0.16 $  \cite{San93}.  We
also obtain the ratio
\be
\frac{\Gamma (B^0 \to D^{*-} \ell^+ \nu_{\ell})}
{\Gamma (B^0 \to D^- \ell^+ \nu_{\ell})}= 2.54,
\ee
that agrees with experimental  result $2.4\pm 0.6$  \cite{PDG96}.
Some  other  model  predictions  for the  $B\to  D^*$  decay  are
collected in Table 4.

\subsection{Decays $B \to \pi \ell \nu_{\ell}$ and
$B \to \rho \ell \nu_{\ell}$}

We now consider the semileptonic  B-meson decay  corresponding to
the quark process $b \to u \ell  \nu_{\ell}$.  The  investigation
of this decay is very important for the  determination of the CKM
matrix element  $|V_{bu}|$, which plays an important role for the
CP  violation  in the  Standard  Model.  The  determination  of $
V_{bu}  $ is one of the most  challenging  measurements  in $ B $
physics.  Very  recently, the CLEO  Collaboration  \cite{Ammar95}
has reported the first signal for exclusive  semileptonic  decays
of the B meson into charmless final states, in particular for the
decay  modes  $B \to  \pi  \ell  \nu_{\ell}$,  $B \to  \rho  \ell
\nu_{\ell}$.  However, there is a significant model dependence in
the simulation of the reconstruction  efficiencies.  The observed
branching ratios, assuming the efficiencies obtained from the ISGW
model \cite{ISGW89} are
  \be
     Br(B \to \pi \ell \nu_{\ell})  =  (1.34 \pm 0.45) \cdot 10^{-4},  ~~~
     Br(B \to \rho \ell \nu_{\ell})  =  (2.28^{+0.69}_{-0.83}) \cdot 10^{-4}.
  \ee
\null From our predicted rates $\Gamma (B^0 \to \pi^- \ell^+ \nu_{\ell}) =
8.72~ |V_{bu}|^2 ~ ps^{-1}$
and $\Gamma (B^0 \to \rho^- \ell^+ \nu_{\ell}) =
13.2~ |V_{bu}|^2 ~ ps^{-1}$ (see Table 2) we obtain
\be
    |V_{bu}| = 0.0031 \pm 0.0004, ~~(B^0 \to \pi^- \ell^+ \nu_{\ell}),
 \ee
\be
    |V_{bu}| = 0.0033 \pm 0.0004,~~ (B^0 \to \rho^- \ell^+ \nu_{\ell}).
\ee
Using our previous result, Eqs.(58, 59), for $|V_{bc}|$ we get
\be
    \left| \frac{V_{bu}}{V_{cb}} \right| = 0.087 \pm 0.015, ~~(B^0 \to \pi^-
\ell^+ \nu_{\ell}),
\ee
 \be
    \left| \frac{V_{bu}}{V_{cb}} \right| = 0.092 \pm 0.015, ~~(B^0 \to \rho^-
\ell^+ \nu_{\ell}).
\ee
These  values  are  in  agreement\footnote   {If  the  BSW  model
\cite{WSB85}  is used to reconstruct  efficiencies  the branching
fractions are somewhat  higher:\\ $ Br(B \to \pi \ell \nu_{\ell})
= (1.63  \pm  0.57)  \cdot  10^{-4}  $, and $ Br(B \to \rho  \ell
\nu_{\ell}) = (3.88^{+1.15}_{-1.39})  \cdot 10^{-4} $.  Combining
the average of  experimental  results for the ISGW and BSW models
with our  predicted  rates one  obtains $  |V_{bu}|  = 0.0033 \pm
0.0004  $, $ \left|  \frac{V_{bu}}{V_{cb}}  \right|  = 0.092  \pm
0.015,  ~~(B^0 \to \pi^-  \ell^+  \nu_{\ell})  $ and $ |V_{bu}| =
0.0039  \pm  0.0004 $, $ \left|  \frac{V_{bu}}{V_{cb}}  \right| =
0.107 \pm 0.015, ~~(B^0 \to \rho^- \ell^+  \nu_{\ell})  $.}  with
the value derived from  measurements  of the end-point  region of
the   lepton   spectrum   in   inclusive    semileptonic   decays
\cite{Albrecht91,Bartelt93} viz.
 \be
    \left| \frac{V_{bu}}{V_{cb}} \right|_{incl} = 0.08 \pm 0.01_{exp}
    \pm 0.02_{th}.
 \ee

A large variety of calculations of the ratio $|V_{bu} / V_{bc}|$,
based  on  various  non-perturbative  approaches,  exists  in the
literature; the results typically lie in the range from $0.06$ to
$0.11$.  A sample of these calculations is given in Table 5.

Note that our result for the ratio of branching fractions
\be
\frac{\Gamma (B^0 \to \rho^- \ell^+ \nu_{\ell})}
{\Gamma (B^0 \to \pi^- \ell^+ \nu_{\ell})}= 1.51
\ee
compares favorably to the recent CLEO result \cite{Ammar95}:
\be
\frac{\Gamma (B^0 \to \rho^- \ell^+ \nu_{\ell})}
{\Gamma (B^0 \to \pi^- \ell^+ \nu_{\ell})}= 1.70^{+0.80}_{-0.30} \pm
0.58^{+0.00}
_{-0.34}
\ee
The dispersion  relation  approach to the ISGW2 model \cite{Me96}
yields for this ratio the values of  $1.17-1.34$.  Our prediction
for the ratio  $\Gamma_L  /  \Gamma_T$  for the $B \to \rho  \ell
\nu_{\ell}$  decay rate is 0.92; other models  yield  $\Gamma_L /
\Gamma_T   =  $  1.13   \cite{Me96},   1.34   \cite{WSB85},   0.3
\cite{IS95}, 0.82 \cite{Jnew}, and $0.52 \pm 0.08$\cite{Ball97}.

\subsection{Decays $D \to K \ell \nu_{\ell}$ and
$D\to K^* \ell \nu_{\ell}$}

In  the  charm   sector,  the  absolute   scale  of   theoretical
predictions for the form factors which  characterize the hadronic
$ D \to K^* $ and $ D \to  \rho $  matrix  elements  of the  weak
currents  can  be  tested,   because  the  CKM  elements  can  be
determined independently of the $D$ semileptonic decay rate using
the  assumption  of CKM  unitarity  and the  smallness of the CKM
matrix  elements  for  $ B  $  decay  \cite{GKR94}.  Semileptonic
decays $ D\to K^*\ell  \nu_{\ell},~\ell=e,\mu $ have been studied
extensively and the form factors have been determined  (with some
assumptions  concerning  their shape).  In  particular,  the CLEO
collaboration   \cite{BEAN93}   has   measured   all  four  $D\to
K(K^*)\ell \nu_{\ell}$ decays in one experiment:
\be
\Gamma (D\to K\ell \nu_{\ell})=(9.1\pm 0.3 \pm 0.6)\cdot 10^{10} s^{-1},
\ee
\be
\Gamma (D\to K^*\l \nu_l)=(5.7\pm 0.7)\cdot 10^{10} s^{-1},
\ee
\be
R=\frac{\Gamma (D\to K^*\l \nu_l)}{\Gamma (D\to K\ell \nu_{\ell})}
=0.62\pm 0.08.
\ee
Using $|V_{cs}|=0.974$ we predict
\be
\Gamma (D\to K\ell \nu_{\ell})=9.38\cdot 10^{10} s^{-1},
\ee
\be
\Gamma (D\to K^*\ell \nu_{\ell})=6.36\cdot 10^{10} s^{-1},
\ee
\be
R=0.68,
\ee
in  an  agreement  with  experimental   data.  Some  other  model
predictions for the $D\to K^*$ decay are collected in Table 6.

\subsection{Decays $D \to \pi \ell \nu_{\ell}$ and $D \to \rho \ell \nu_{\ell}$
}
Assuming  $|V_{cd}|  =  0.221  $,  which  is  inferred  from  the
unitarity  of the CKM matrix  \cite{PDG96},  we  predict  for the
semileptonic decay rates the following values:

 \be
     \Gamma (D^0 \to \pi^- e^+ \nu_e) = 7.4 \cdot 10^{-3} ~ ps^{-1},
 \ee
\be
     \Gamma (D^0 \to \rho^- e^+ \nu_e) = 3.8 \cdot 10^{-3} ~ ps^{-1}.
 \ee
These predictions should be compared with the experimental result
\be
   \label{3.13}
   \Gamma_{exp} (D^0 \to \pi^- e^+ \nu_e) = (9.4^{+5.5}_{-2.9}) \cdot
   10^{-3} ~ ps^{-1},
 \ee
and the recent light-front QCD sum rule prediction \cite{Khodjamirian95}.
\be
   \label{3.14}
   \Gamma_{SR} (D^0 \to \pi^- e^+ \nu_e) = (7.6 \pm 0.2) \cdot 10^{-3}
   ~ ps^{-1}.
\ee

The CLEO collaboration \cite{Butler95}
recently determined the following ratio
of decay rates
\be
\frac{\Gamma (D^0 \to \pi^- e^+ \nu_{e})}
{\Gamma (D^0 \to  K^{-} e^+ \nu_{e})} = 0.103 \pm 0.039 \pm 0.013,
\ee
while the $E653$ collaboration \cite{Kodama93} measured
\be
 \frac{\Gamma (D^0 \to \rho^- e^+ \nu_{e})}
{\Gamma (D^0 \to K^{*-} e^+ \nu_{e})} = 0.088 \pm 0.062 \pm 0.028 .
\ee
Our calculated fractions are 0.079 and 0.060, respectively, which
seem to be a bit small;  nevertheless  they are in agreement with
the  experimental   values  within  the  large  error  bars.  The
dispersion  approach to the ISGW2  model  \cite{Me96}  yields for
these fractions $0.071$ and $0.047-0.056$.

\section{Conclusions}

In this paper we have examined the weak  transition  form factors
which govern the heavy-to-heavy  and heavy-to-light  semileptonic
decays of pseudoscalar  mesons within a relativistic  constituent
quark model based on the light-front formalism.  We have obtained
all the relevant  expressions  for the weak vector and axial form
factors.  For  numerical   estimates  we  employed  the  LF  wave
function, that is related to the equal-time  wave function of the
ISGW model.  The transition  form factors have been calculated in
the whole kinematical  region accessible in semileptonic  decays.
We have  calculated  the form factors and the decay rates for the
$B  \to  D(D^*)  \ell   \nu_{\ell}$,   $B  \to  \pi  (\rho)  \ell
\nu_{\ell}$, $D \to K(K^*) \ell  \nu_{\ell}$ and $D \to \pi (\rho
) \ell  \nu_{\ell}$  semileptonic  decays.  Our results have been
successfully  compared  with  available   experimental  data.  In
particular  for $b \to c$ and $b \to u$  transitions,  using  the
available  experimental  information  on the  semileptonic  decay
rates, the CKM parameters have been estimated.

Since  our  calculations  neglect  the  contribution  of the pair
creation from the vacuum it would appear that the good  agreement
with the data  suggests  that  these  might  very well be  small.
However, an estimate of such contributions,  particularly in case
of the heavy-to-light transitions, is important for there to be a
complete comparison with experimental data.

\section*{Acknowledgment}
We thank I.L.  Grach for the collaboration in the early stages of
this work.  The authors  acknowledge the financial support of the
INTAS--RFBR  grant,  ref.  No  95--1300.  This  work  was in part
supported  by the RFBR grant,  ref.  No  95-02-04808a  and by the
Natural Sciences and Engineering Council of Canada.

\setcounter{equation}{0}
\def\theequation{A.\arabic{equation}}

\section*{Appendix. Form factors at $q^2=0$}

In this  Appendix  we  show  that  Eqs.  (53-55)  for $ V(0) $, $
A_0(0) $ and $ A_2(0) $ coincide with the ones obtained using the
LF  formalism  in the  Breit  frame  for $ q^2\le 0 $.  Using our
notation we  write\footnote{Note  that our variables and notation
are related to those of ref.  \cite{Jold} by
$$
{\bf p_{\bot}''}=-{\bf k_{\bot}'},~~{\bf p_{\bot}'}=-{\bf k_{\bot}},~~\xi=1-x,
~~m_2=m,~~m_1''=m_2,~~m_1'=m_1,
$$
The quantity $\Omega (p'',p') $ defined by Eq.(4.2) of \cite{Jold} can
be written in terms of our LF functions as
$$
\frac{dk_3}{dx} \frac{1}{1-x} \Omega (k',k)=\frac{1}{x}
\chi_2 (x,k'^2_{\bot}) \chi_1 (x,k^2_{\bot})
$$}
Eqs.(4.4, 4.5, 4.6a, 4.9) of Ref. \cite{Jold} in the form
\be
V(q^2)=\frac{M_1+M_2}{\sqrt2M_1}\int\limits^1_0 \frac{dx}{x}\int
d^2k_{\bot}\chi_2(x,k'^2_{\bot})
\chi_1(x,k^2_{\bot})
{\cal I}^{(+1)}_V(x;k_{\bot};q_{\bot}),
\ee
\be
A_0(q^2)=\frac{1}{2M_1}\int\limits^1_0\frac{dx}{x}\int d^2k_{\bot}
\chi_2(x,k'^2_{\bot})\chi_1(x,k^2_{\bot})
{\cal I}^{(0)}_A(x;k_{\bot};q_{\bot}),
\ee
\be
A_2(q^2)=\frac{M_1+M_2}{\sqrt2M_1}\int\limits^1_0\frac{dx}{x}\int d^2k_{\bot}
\chi_2(x,k'^2_{\bot})\chi_1(x,k^2_{\bot})
{\cal I}^{(+1)}_A(x;k_{\bot};q_{\bot}),
\ee
where
\be
{\cal I}^{(+1)}_V(x;k_{\bot};q_{\bot})=
\sqrt2M_1\left[{\cal A}_1
 -\frac{m_1-m_2}{q^2}{\bf k_{\bot}q_{\bot}}+
\frac{2}{M_{20}+m_2+m}
\left[k^2_{\bot}+\frac{({\bf k_{\bot}q_{\bot}})^2}{q^2}\right]\right],
\ee
\be
{\cal I}^{(0)}_A(x;k_{\bot};q_{\bot})=
2M_1\left[2(1-x) M_{20}{\cal A}_1+
 \frac{(1-2x)M_{20}+m_2-m}{x(M_{20}+m_2+m)}
({\bf k_{\bot}k_{\bot}'}+{\cal A}_1{\cal B})\right],
\ee
$$
{\cal I}^{(+1)}_A(x;k_{\bot};q_{\bot})=
-\sqrt2M_1\left[(2x-1){\cal A}_1 -\right.
$$
\be
\left. \frac{{\bf k_{\bot}q_{\bot}}}{q^2}[2(1-x) m+m_2+(2x-1)m_1] +
2\frac{1-\frac{{\bf k_{\bot}q_{\bot}}}{xq^2}}
{M_{20}+m_2+m}
({\bf k_{\bot}k_{\bot}'}+{\cal A}_2{\cal B})\right],
\ee
and $ {\bf k'_{\bot}}={\bf k_{\bot}}+x{\bf q_{\bot}}$ .

Noting  that  in  our   kinematics   the  momentum   transfer  is
anti-parallel to the 1-axis, i.e.,
$ q=(q^0,q^1,q^2,q^3)=(0,-q_{\bot},0,0) $
we obtain after simple algebra
\be
{\cal I}^{(+1)}_V(x;k_{\bot};q_{\bot})=
\frac{1}{q_{\bot}}I^{(+1)}_V(x,1,k_{\bot},q_{\bot}),
\ee
\be
{\cal I}^{(0)}_A(x;k_{\bot};q_{\bot})=
I^{(0)}_A(x;1;k_{\bot};q_{\bot}),
\ee
\be
{\cal I}^{(+1)}_A(x;k_{\bot};q_{\bot})=
\frac{1}{q_{\bot}}I^{(+1)}_A(x;1;k_{\bot};q_{\bot}),
\ee
In the  limit $  q_{\bot}\to  0 $  using  Eqs.  (33,  35, 36) one
easily  derives  that Eqs.  (A.1,  A.2, A.3)  coincide  with Eqs.
(53, 54, 55).

\newpage

\bigskip

\vspace{1cm}

{\bf Table 1}.  Form  factors at $q^2=0$  and decay rates for the
set of quark model  parameters  of  \cite{Jnew}.  The decay rates
for   $bc$   and   $bu$    transitions    are   in   the    units
$|V_{bc}|^2ps^{-1}$ and  $|V_{bu}|^2ps^{-1}$,  respectively.  The
decay  rates  for  $cs$  and  $cd$   transitions   are  in  units
$10^{10}s^{-1}$.   The   values   of   the   CKM   parameters   $
|V_{cd}|=0.22  $, $  |V_{cs}|=0.9743  $ are used.  In parentheses
are given the numerical results of ref.  \cite{Jnew}.

\vspace{0.5cm}

\medskip

\bt{|l|l|l|l|l|l|l|}\hline
$Q_1Q_2$&$F_1(0)$&$V(0)$&$A_1(0)$&$A_2(0)$&$\Gamma (PS\to PS)$&
$\Gamma (PS\to V)$\\ \hline
bc&$0.68(0.69)$&$0.79(0.81)$&$0.67(0.69)$&
$0.61(0.64)$&$9.3(9.6)$&
$24.5(25.3)$\\
bu&$0.25(0.27)$&$0.34(0.35)$&$0.26(0.26)$&$0.24(0.24)$&
$8.1(10)$&$18.3(19.1)$\\
cs&$0.78(0.78)$&$1.03(1.04)$&$0.65(0.66)$&$0.43(0.43)$&
$9.7(9.6)$&$6.2(5.5)$\\
cd&$0.66(0.67)$&$0.90(0.93)$&$0.57(0.58)$&$0.42(0.42)$&
$0.7(0.8)$
&$0.39(0.33)$\\ \hline
\et

\vspace{0.5cm}

{\bf Table 2}.  Form  factors at $q^2=0$  and decay rates for the
set of quark model  parameters  of  \cite{IS95}.  The decay rates
for   $bc$   and   $bu$    transitions    are   in   the    units
$|V_{bc}|^2ps^{-1}$ and  $|V_{bu}|^2ps^{-1}$,  respectively.  The
decay   rates   for   $cs$   and   $cu$    transitions   are   in
$|V_{cs}|^210^{10}s^{-1}$     and      $|V_{cd}|^210^{10}s^{-1}$,
respectively.  In  parentheses  are given the decay rates for the
$PS-PS$ transitions taken from ref.  \cite{DGNS96}.

\vspace{0.5cm}

\medskip

\bt{|l|l|l|l|l|l|l|l|}\hline
$Q_1Q_2$&$F_1(0)$&$V(0)$&$A_1(0)$&$A_2(0)$&$A_0(0)$&$\Gamma (PS\to PS)$&
$\Gamma (PS\to V)$\\ \hline
bc&$0.683$&$0.677$&$0.623$&
$0.556$&$0.678$&$9.09(9.78)$&
$23.10$\\
bu&$0.293$&$0.216$&$0.170$&$0.155$&
$0.214$&$8.72(9.62)$&$13.2$\\
cs&$0.780$&$0.777$&$0.633$&$0.464$&
$0.73$&$9.88(11.30)$&$6.70$\\
cd&$0.681$&$0.663$&$0.502$&$0.366$&
$0.60$&$15.29(16.0)$
&$7.85$\\ \hline
\et

\vspace{0.5cm}

{\bf Table 3}.  Model-dependent predictions of the $\bar B\to D^*
\ell \bar \nu $ partial width  $V_{cb}$  values  derived from the
measured partial width.  An additional  $3.5\%$  systematic error
due to $B$ lifetime  measurements is common to all the $|V_{cb}|$
values given.

\vspace{0.5cm}

\medskip

\noindent
\bt{|l|l|l|l|l|}\hline
&This work&ISGW\cite{ISGW89}&BSW\cite{WSB85}&KS\cite{KS88}$$\\ \hline
$\Gamma(\bar B\to D^*\ell \bar \nu)|V_{cb}|^{-2}$&$23.1ps^{-1}$&
$24.6ps^{-1}$&$21.9ps^{-1}$&$25.8ps^{-1}$\\ \hline
$10^3\cdot |V_{cb}|$&$35.6\pm 1.1\pm 1.6$&$34.8\pm 1.1\pm1.6$&
$37.5\pm 1.2\pm 1.7$&$34.4\pm 1.1\pm 1.5$\\ \hline
\et

\vspace{0.5cm}

{\bf Table 4}.  Form  factors at $q^2=0$ and the decay  widths in
for  the   semileptonic   $B^0\to  D^*$   transition.  The  model
predictions  for  the  width  are  in  $|V_{bc}|^2~ps^{-1}$.

\vspace{0.5cm}

\medskip

\noindent
\bt{|l|l|l|l|l|l|l|}\hline
$B\to D,D^*$& $V(0)$& $V(0)/A_1(0)$&
$A_2(0)/A_1(0)$& $\Gamma(D^*)$& $\Gamma(D)/\Gamma(D^*)$
&$\Gamma_L/\Gamma_T$\\
\hline
This
work&$0.68$&$1.08$&$0.89$&$23.1$&$0.38$&$1.15$\\
DR\cite{Me96}&0.68&1.08&0.89&21.0&0.41&$1.28$\\
ISGW2\cite{IS95}&&&&24.8&0.48&$1.04$\\
KS\cite{KS88}&&1.54&1.39&25.8&0.32&$$\\
BSW\cite{WSB85}&0.71&1.09&1.06&22.0&0.36&$$\\
Stech \cite{Stech96}&$0.75$&$1.10$&$1.02$&$24$&$0.37$&$$\\
SR \cite{Narison92}&$0.58$&$1.26\pm0.08$&
$1.15\pm0.20$&$17\pm6$&$0.53\pm0.11$&$$\\
Exp. \cite{Gronberg95},\cite{San93}  &&$1.18\pm0.30\pm0.12$
&$0.71\pm0.22\pm0.07$&&$0.46\pm0.2$&$1.24\pm0.16$\\
\hline
\et

\newpage

\bigskip

{\bf Table 5}.  Values  for  $|\frac{V_{ub}}{V_{cb}}|$  extracted
from CLEO  measurement  of exclusive  semileptonic  B decays into
charmless final states.  Following  Neubert  \cite{Neubert96}  we
take in this  comparison  $|V_{cb}|=0.040$.  An average  over the
experimental  results  obtained  using the ISGW and BSW models to
reconstruct  efficiencies is used for all except the ISGW and BSW
models, where the numbers corresponding to these models are used.
For our LF model we use the values  quoted in  Eq.(64,  65).  The
first error quoted is experimental,  the second (when  available)
is theoretical.  For the recent LC SR result  \cite{Ball97}  only
the theoretical error is quoted.

\vspace{0.5cm}

\medskip

\bt{|l|l|l|l|}\hline
Method&Reference&$\bar B\to \pi \ell \bar \nu$
&$\bar B\to \rho \ell \bar \nu$\\
\hline
LF QM & this work &$ 0.078\pm 0.013 $&$ 0.083^{+0.013}_{-0.015}$\\
\hline
QCD sum rules&Narison \cite{Narison92}&$0.159\pm 0.019\pm 0.001$&
$0.066^{+0.007}_{-0.009}\pm 0.003$\\
&Ball \cite{Ball97}&$$&$0.086\pm0.013$ \\
&KhR \cite{Khodjamirian95}&
$0.085\pm 0.010$&$$\\
\hline
Lattice QCD & UKQCD\cite{UKQCD95}&$0.103\pm 0.012^{+0.012}_{-0.010}$&$$\\
&APE \cite{APE95}&$0.084\pm 0.010\pm 0.021$&$$ \\
\hline
pQCD& Li and Yu \cite{LY95} &$ 0.054\pm 0.006 $&$$ \\
\hline
Quark models&BSW \cite{WSB85} &$ 0.093\pm 0.016 $&$0.076^{+0.011}_{-0.014}$\\
&KS \cite{KS88} &$ 0.088\pm 0.011$&$ 0.056^{+0.006}_{-0.007} $\\
&ISGW2 \cite{IS95}&$ 0.074\pm 0.012 $&$ 0.079^{+0.012}_{-0.014} $\\
\hline
\et

\vspace{1cm}

{\bf Table 6}. Form factors at $q^2=0$ and the decay widths in
$10^{10}~s^{-1}$ for the semileptonic $D^0\to K^*$ transition.
For our calculations we use $|V_{cs}|=0.974$

\vspace{0.5cm}

\medskip

\noindent
\bt{|l|l|l|l|l|l|}\hline
$D^0\to K^*$&$V(0)$&$A_1(0)$&$A_2(0)$&$\Gamma(K^*)$&
$\Gamma_L/\Gamma_T $\\ \hline
This work&$0.777$&$0.633$&$0.464$&$6.36$&$1.30$\\
DR \cite{Me96}&0.777&0.633&0.464&5.38&1.31\\
SR \cite{BBD91}&
$1.1\pm0.25$&$0.50\pm0.15$&$0.60\pm0.15$&$3.8\pm1.5$&$0.86\pm0.06$\\
Lat. \cite{APE95}&
$1.08\pm0.22$&$0.67\pm0.11$&$0.49\pm0.34$&$6.9\pm1.8$&$1.2\pm0.3$\\
Lat. \cite{UKQCD95}&
$1.01^{+0.3}_{-0.13}$&$0.70^{+0.07}_{-0.10}$&$0.60^{+0.10}_{-0.15}$&
$6.0^{+0.8}_{-1.6}$&$1.06\pm0.16$\\
Stech \cite{Stech96}&$1.07$&$0.69$&$0.73$&$7.1$&$0.97$\\
Exp. \cite{PDG96}&$1.1\pm0.2$&$0.56\pm0.04$&$0.40\pm0.08$&$5.7\pm0.7$&
$1.15\pm0.17$\\ \hline
\et

\newpage

\section*{\bf Figure Captions}
\begin{description}

\item[Figures  1, 2, 3 and 4] The form factors $ F_1(q^2)$  for $
B\to  D $,  $B\to  \pi  $,  $D\to  K $  and  $  D\to  \pi  $
transitions,  respectively,  obtained using the ISGW  parameters,
but adopting  the  experimental  values for the meson masses.  The
relation between the kinematical variables $y$ and $q^2$ is given
by  Eq.(4).  The  solid   lines  are  the   results   of  our  LF
calculations  obtained using Eq.(46) with $\alpha_1, \alpha_2 \to
0$.  The  dashed   lines  are  the   previous   results  of  ref.
\cite{DGNS96}  corresponding  to the choice of the two  reference
frames with $\alpha_1 =0$,  $\alpha_2 = \pi $ as explained in the
text.  The corresponding decay rates are listed in Table 2.

\item[Figures 5, 6, 7 and 8] The vector $ V(q^2) $ and axial form
factors  $A_0(q^2)$,  $A_1(q^2)$ and  $A_2(q^2)$  for $B\to D^*$,
$B\to  \rho  $,  $D\to  K^*$  and  $D  \to  \rho  $  transitions,
respectively,  obtained using the ISGW  parameters,  but adopting
the   experimental   values  for  the  meson  masses.  The  solid,
dashed,  dotted  and  dot-dashed  lines  represent the
vector form factor  $V(q^2)$  and axial form factors  $A_0(q^2)$,
$A_1(q^2)$ and $A_2(q^2)$, respectively.  The corresponding decay
rates are listed in Table 2.

\item[Figure 9] The  $\frac{d\Gamma}{dq^2}  $ distribution  for $
\bar B\to D^*\ell  \nu_{\ell}$  decays.  The upper thin line with
$|V_{bc}|=0.0374  $  corresponds to  our  fit  to the  CLEO  data
\cite{Barish95}  while  the  lower  solid  line corresponds  to $
|V_{bc}|=0.036 $ found from the partial width.

\end{description}

\newpage
\begin{figure}[htb]
\vskip 8.5in\relax\noindent\hskip -.5in\relax{\includegraphics{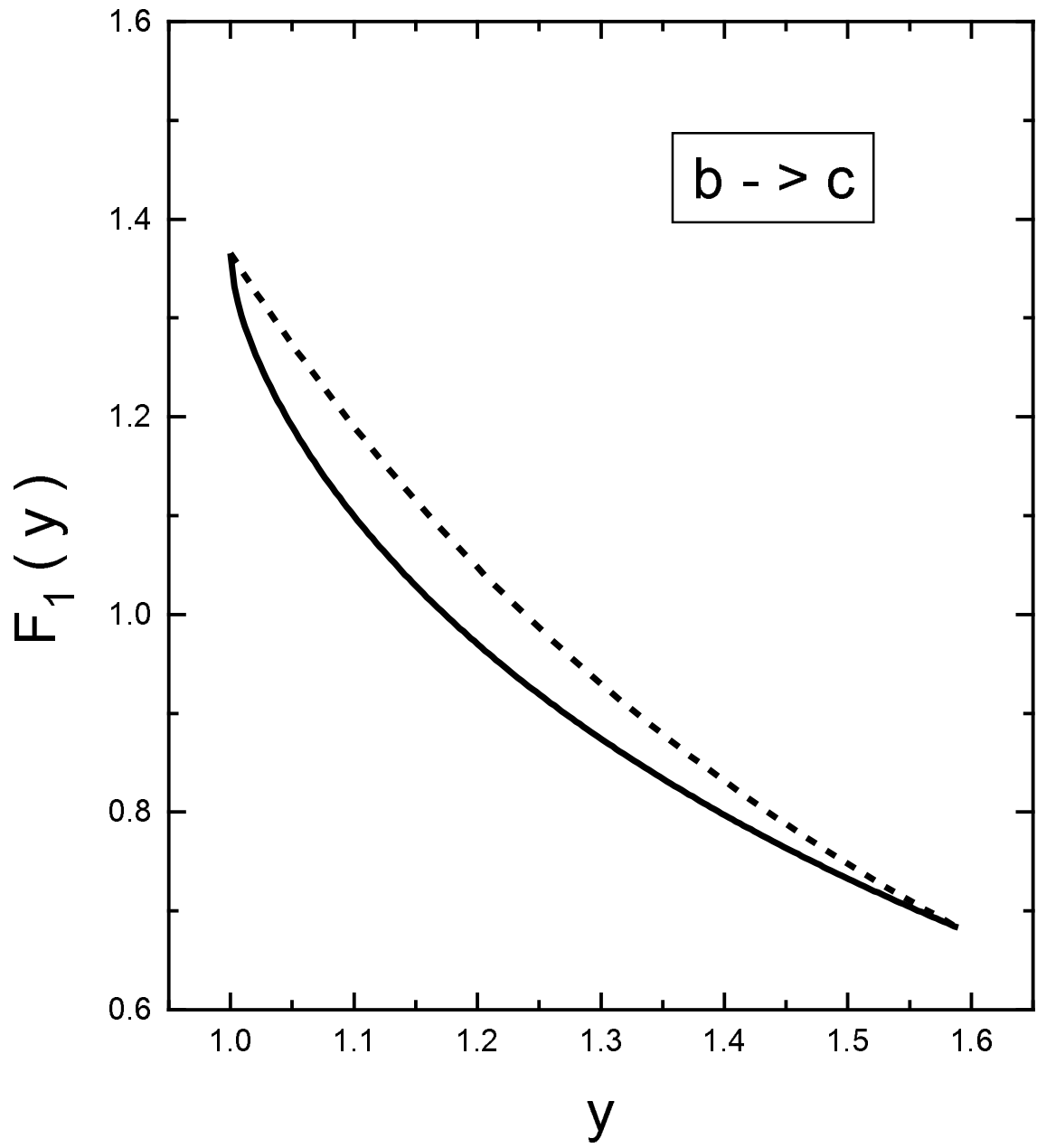}}
\caption{Form factor $F_{1}(q^2)$ for $B \to D$}
 \end{figure}
 \newpage
 \begin{figure}[htb]
 \vskip 8.5in\relax\noindent\hskip -.5in\relax{\includegraphics{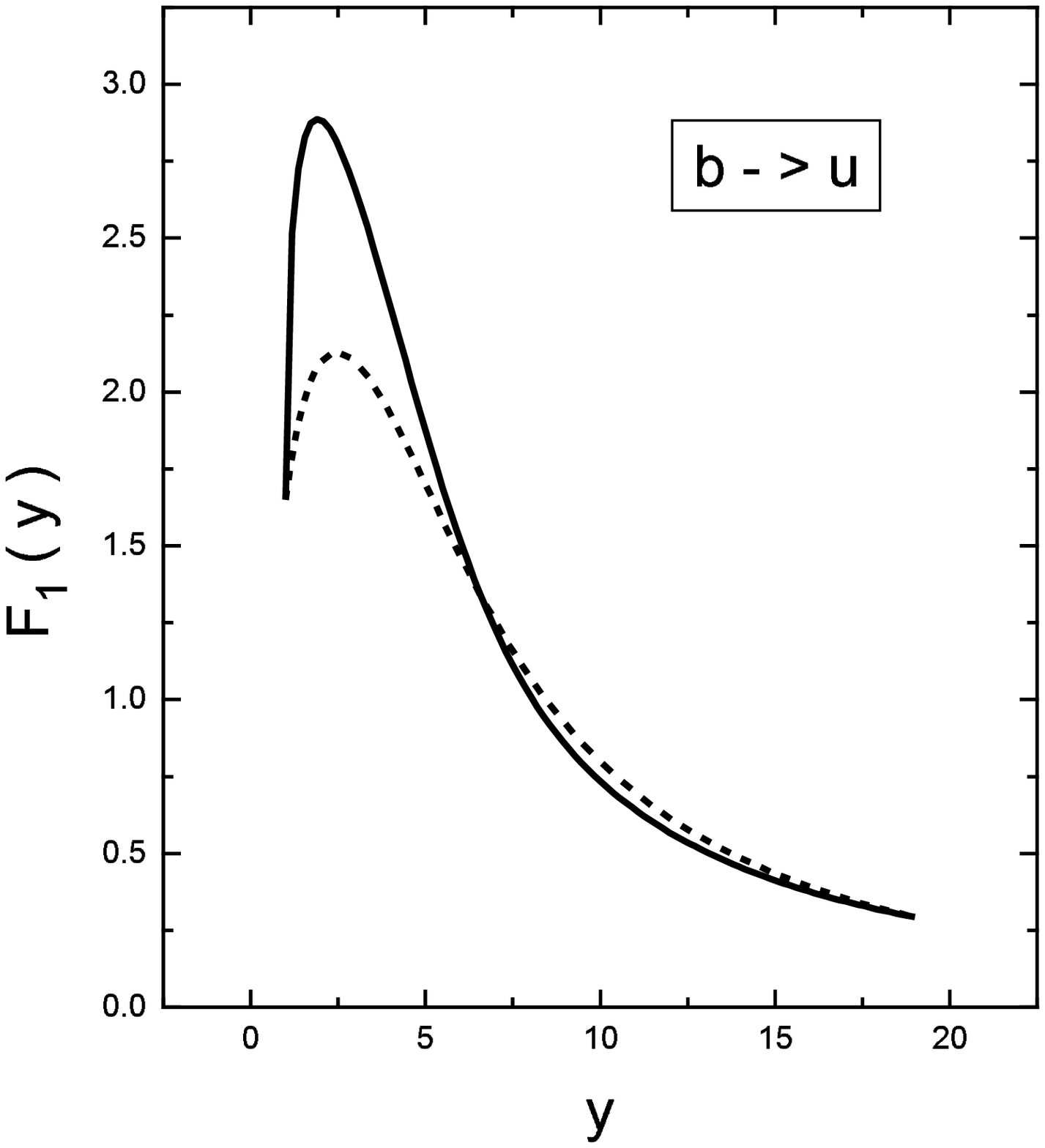}}
 \caption{Form factor $F_{1}(q^2)$ for $B \to \pi$}
 \end{figure}
 \newpage
 \begin{figure}[htb]
 \vskip 8.5in\relax\noindent\hskip -.5in\relax{\includegraphics{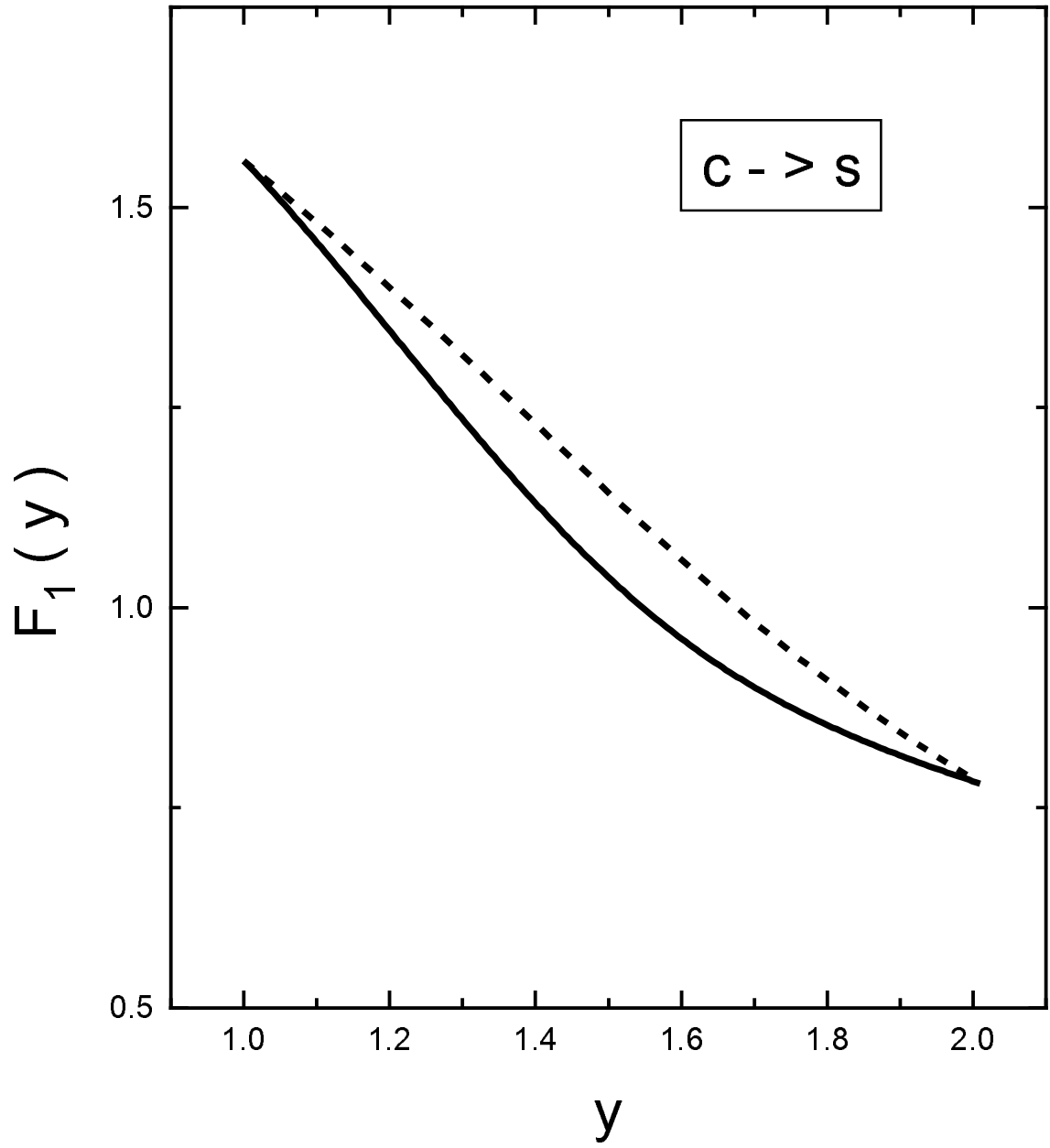}}
 \caption{Form factor $F_{1}(q^2)$ for $D \to K$}
 \end{figure}
 \newpage
 \begin{figure}[htb]
 \vskip 8.5in\relax\noindent\hskip -.5in\relax{\includegraphics{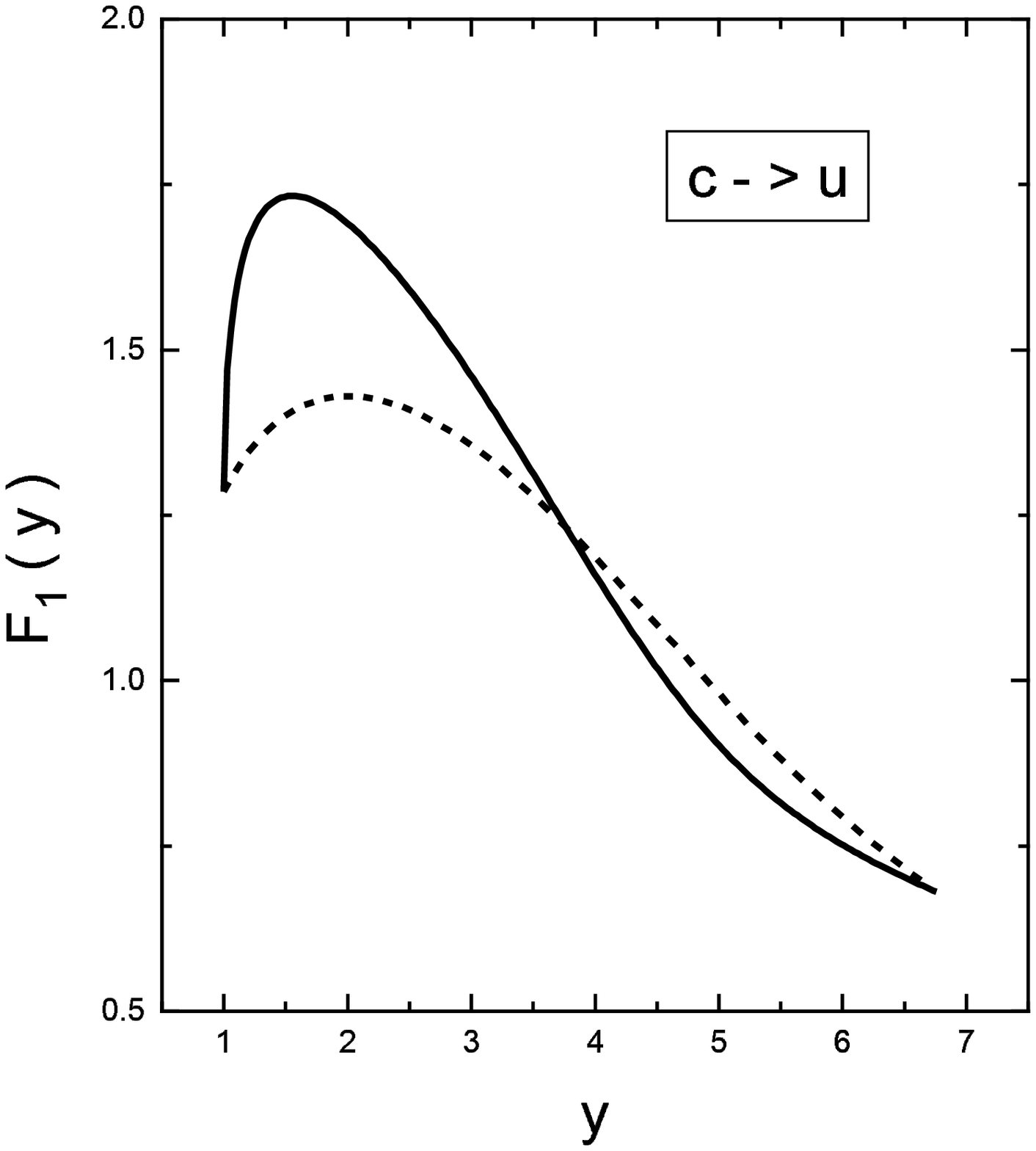}}
 \caption{Form factor $F_{1}(q^2)$ for $D \to \pi$}
 \end{figure}
 \newpage
 \begin{figure}[htb]
 \vskip 8.5in\relax\noindent\hskip -.5in\relax{\includegraphics{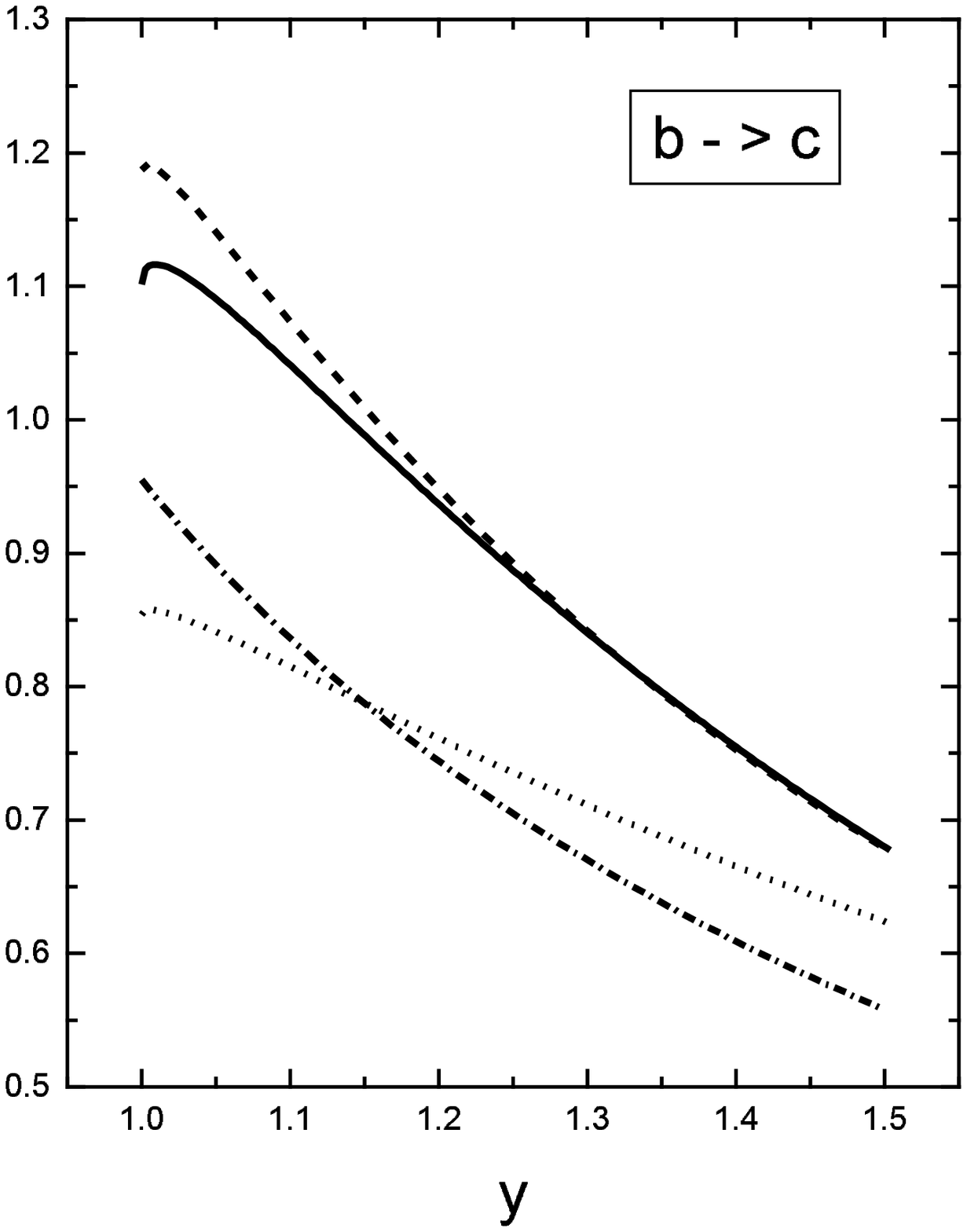}}
 \caption{The vector $ V(q^2) $ and axial form factors $A_0(q^2)$,
 $A_1(q^2)$ and $A_2(q^2)$ for $B\to D^*$.}
 \end{figure}
 \newpage
 \begin{figure}[htb]
 \vskip 8.5in\relax\noindent\hskip -.5in\relax{\includegraphics{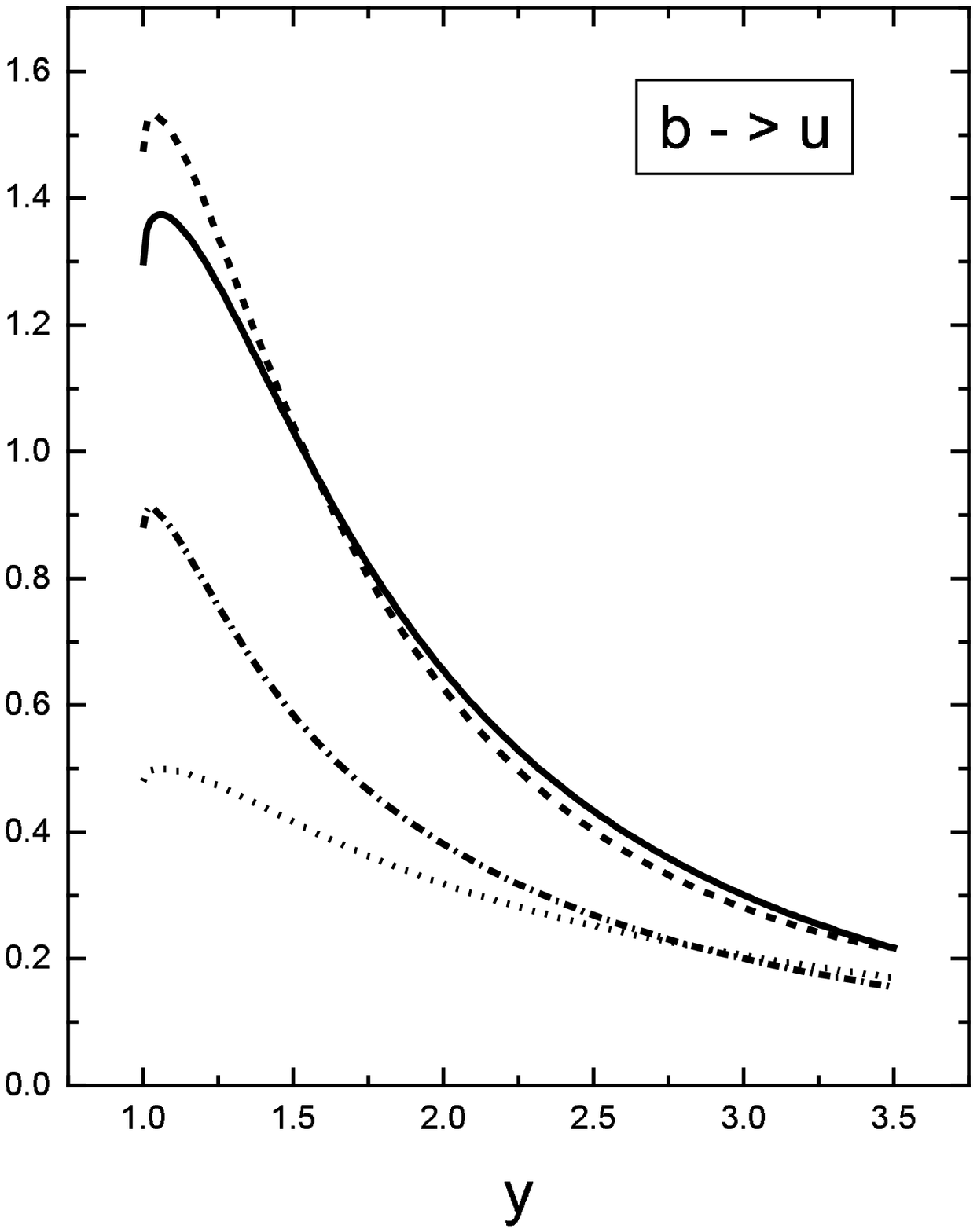}}
 \caption{ The vector $ V(q^2) $ and axial form factors $A_0(q^2)$,
 $A_1(q^2)$ and $A_2(q^2)$ for  $B\to \rho $.  }
 \end{figure}
 \newpage
 \begin{figure}[htb]
 \vskip 8.5in\relax\noindent\hskip -.5in\relax{\includegraphics{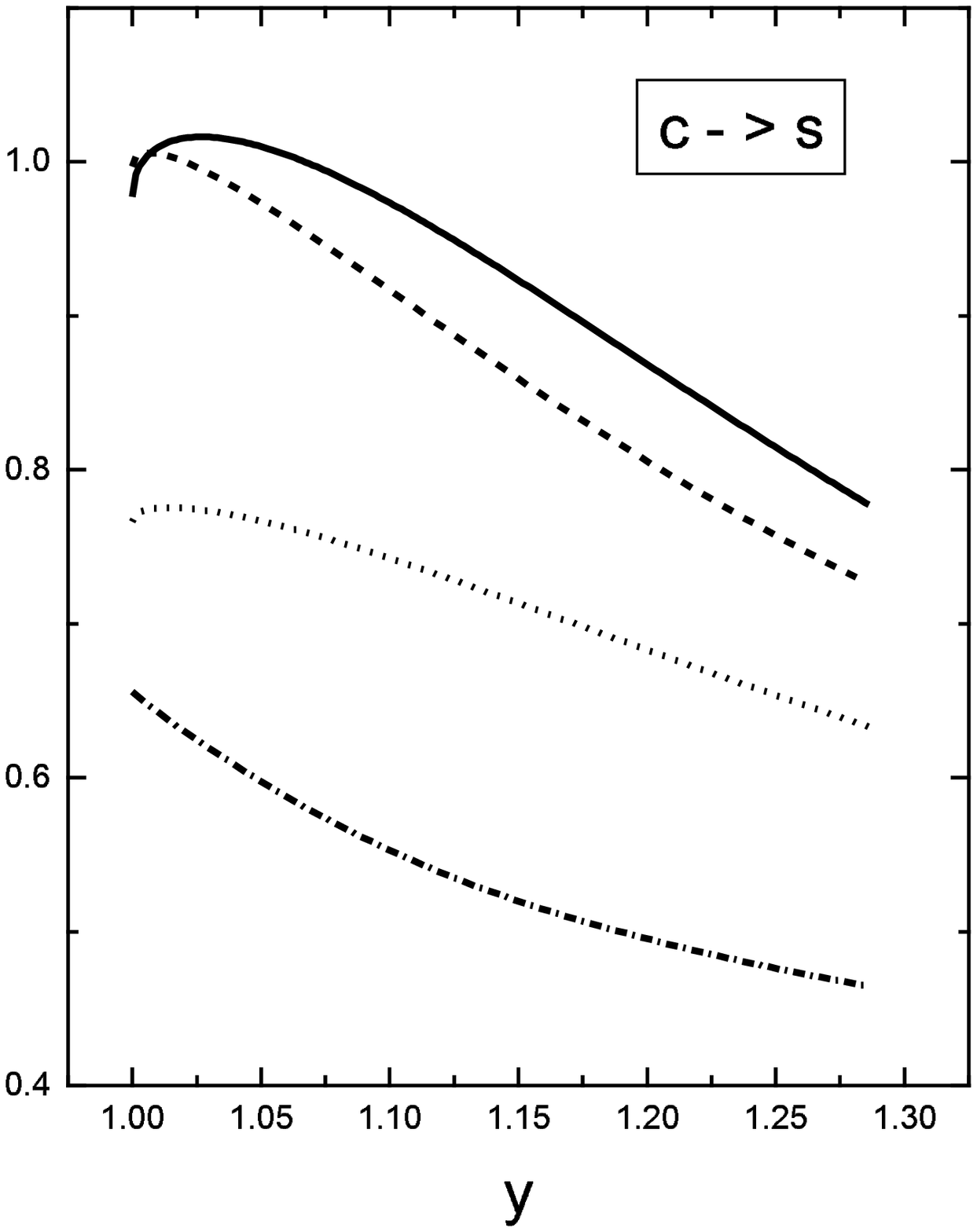}}
 \caption{ The vector $ V(q^2) $ and axial form factors $A_0(q^2)$,
 $A_1(q^2)$ and $A_2(q^2)$ for $D\to K^*$.  }
 \end{figure}
 \newpage
 \begin{figure}[htb]
 \vskip 8.5in\relax\noindent\hskip -.5in\relax{\includegraphics{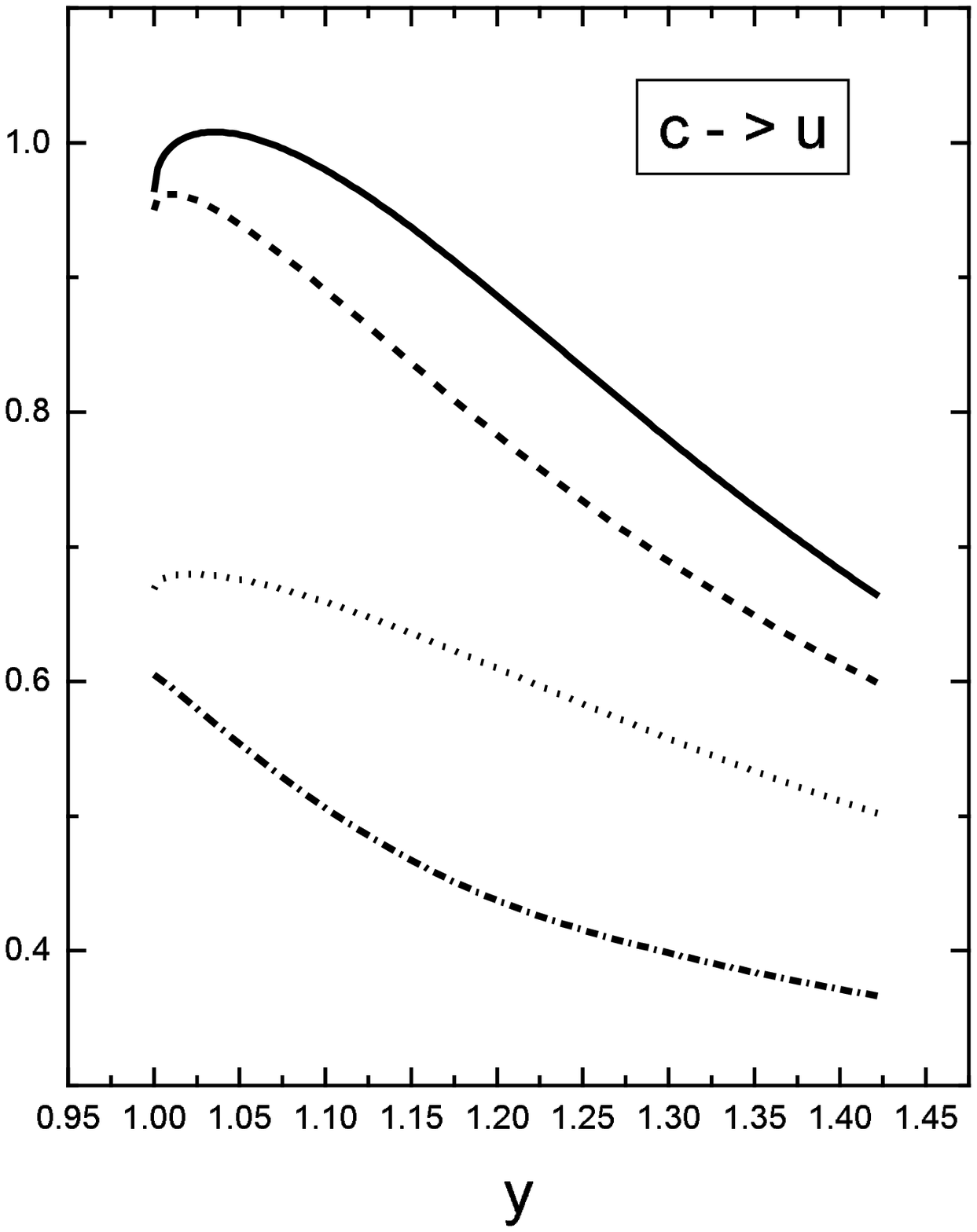}}
 \caption{
 The vector $ V(q^2) $ and axial form factors $A_0(q^2)$,
 $A_1(q^2)$ and $A_2(q^2)$ for $D\to \rho$.  }
 \end{figure}
\newpage
\begin{figure}[htb]
\vskip 8.5in\relax\noindent\hskip -.5in\relax{\includegraphics{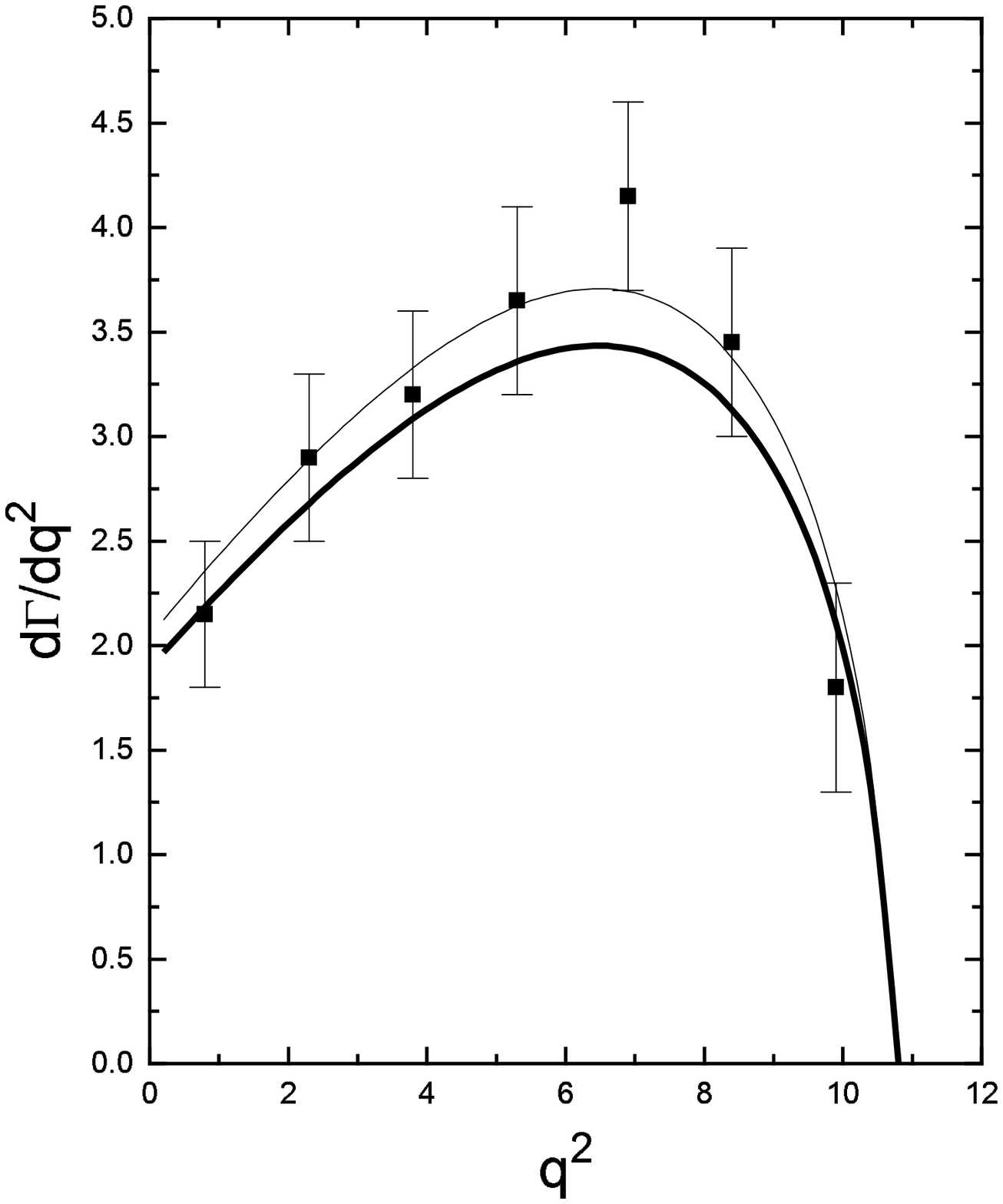}}
\caption{ The $\frac{d\Gamma}{dq^2} $ distribution for
$ \bar B\to D^*\ell \nu_{\ell}$ decays. }
\end{figure}

\newpage


\begin{thebibliography}{99}


\bibitem{NEU94} For a review see M.Neubert, Phys. Rep. {\bf 245} (1994) 259.

\bibitem{QM} For the literature on quark models see\\
D.Melikhov, Phys. Rev. {\bf D53} (1996) 2460; Phys. Lett. {\bf 380} (1996)
363;\\
R.N.Faustov et al., Phys. Rev. {\bf D53} (1996) 6302;\\
M.A.Ivanov and Y.M.Valit, hep--ph/9606404;\\
H.Y.Cheng {\it et al.}, hep--ph/9607332.

\bibitem{DGNS96} N.B.Demchuk, I.L.Grach, I.M.Narodetskii and S.Simula,
Yad. Phys. {\bf 59} (1996) 2235.

\bibitem{GNS96} I.L.Grach, I.M.Narodetskii and S.Simula, Phys.Lett.
{\bf B385} (1996) 317.

\bibitem{TB} M.V. Terent'ev, Yad.Fiz. {\bf 24} 207
[Sov. J. Nucl. Phys. {\bf 24} (1976) 106];
V.B. Berestetsky and M.V. Terent'ev, Yad.Fiz. {\bf 24} (1976) 1044
[Sov. J. Nucl. Phys {\bf 24} (1976) 547].

\bibitem{DF93} C.O. Dib and F. Fera, Phys. Rev. {\bf D47} (1993) 3938.

\bibitem{DT96} P.J.O'Donnell and G.Turan, UTPT-96-6.

\bibitem{Jold} W.Jaus,  Phys. Rev. {\bf D41} (1990) 3394.

\bibitem{DXT95} P.J.O'Donnell, Q.P.Xu and H.K.K.Tung, Phys. Rev.
{\bf D52} (1995) 3966.

\bibitem{Jnew} W.Jaus,  Phys. Rev. {\bf D53} (1996) 1349, and private
communication.

\bibitem{IS95}  D.Scora, N.Isgur,  Phys. Rev. {\bf D52} (1995) 2783.

\bibitem{Gr96}  I.L.Grach, private communication.

\bibitem{PDG96} Particle Data Group, R.M.Barnet {\it et al}, Phys.
Rev. {\bf D53} (1996) 1.

\bibitem{IW90} N.Isgur, M. Wise, Phys. Rev {\bf D41} (1990) 151.

\bibitem{Richman96} J.D.Richman, ref.\cite{PDG96}, p.482.

\bibitem{Barish95} CLEO Collaboration: B.Barish {\it et al}.,
Phys. Rev. {\bf D51} (1995) 1014.

\bibitem{VENUS94} W.Venus, in {\it Lepton and Photon Interactions}, Proceedings
of the XVI International Symposium, Ithaca, New York, 1993, edited by P.Drell
and D.Rubin, AIP Conf. Proc. No 302 (AIP, New York, 1994).

\bibitem{Gronberg95} J.Gronberg {\it et al.} CLEO preprint CLEO--CONF--95--27
(1995).

\bibitem{Skwar95} T.Skwarnicki, in Proceedings of the $17^{th}$ International
Symposium on Lepton-Photon Interactions, Beijing (China), August 1995,
eds. Z. Zhi-Peng and C. He-Sheng (World Scientific, Singapore, 1996), p.238

\bibitem{Ammar95} R.Ammar et al. CLEO-CNF 95-09 (1995) [eps0165].

\bibitem{Me96} D.Melikhov, hep-ph/9611364.

\bibitem{ISGW89} N. Isgur, G. Scora, B. Grinstein, M. Wise, Phys. Rev.
{\bf D39} (1989) 799.

\bibitem{WSB85} M.Wirbel, B.Stech, and  M.Bauer, Z. Phys. {\bf C29} (1985) 637.

\bibitem{Albrecht91} ARGUS Collaboration, H. Albrecht {\it et al.}
Phys. Lett. {\bf B255} (1991) 297.

\bibitem{Bartelt93} CLEO Collaboration, J.Bartelt {\it et al.} Phys. Rev.
Lett.  {\bf 71} (1993) 4111.

\bibitem{Khodjamirian95} A.Khodjamirian and R.R\"{u}ckl Nucl. Instr. and Meth.
{\bf A368} (1995) 28.

\bibitem{Butler95} CLEO Collaboration, F.Butler {\it et al.} Phys. Rev.
{\bf D52} (1995) 2656.

\bibitem{Kodama93} E653 Collaboration, K.Kodama {\it et al.} Phys. Lett.
{\bf B316} (1993) 455.

\bibitem{BEAN93} A.Bean {\it et al}., Phys. Lett. {\bf B317}, (1993) 647.

\bibitem{GKR94} F.J.Gilman {\it et al}. ref. \cite{PDG96} 94.

\bibitem{Ball97} P.Ball, V.M.Braun, hep--ph/9701238, and private communication

\bibitem{Stech96} B.Stech, Preprint HD--THEP--96--35.

\bibitem{San93} S.Sanghera {\it et al.}, Phys.Rev. {\bf D47} (1993) 791.


\bibitem{Neubert96} M.Neubert, {\it B Decays and CP Violation},
CERN-TH/96-55, to appear in International Journal of Modern Physics A.

\bibitem{Narison92} S.Narison   Phys. Lett. {\bf B283} (1992) 384.

\bibitem{BBD91} P. Ball, V.M. Braun, H.G. Dosch, Phys. Rev. {\bf D44}
(1991) 3567. 

\bibitem{UKQCD95} UKQCD collaboration, D.R. Burford et al.,
Nucl. Phys. {\bf B 447} (1995) 425.

\bibitem{APE95} APE collaboration: C.R.Allton et al.,
Phys. Lett. {\bf B345} (1995) 513.

\bibitem{LY95} H-n. Li and H.-L. Yu, Phys. Rev. Lett. {\bf 74}, (1995) 4388.


\bibitem{KS88} J.G. K\"{o}rner and G.A. Schuler,
Z. Phys. {\bf C 38}, (1988) 511.

\end{thebibliography}
\end{document}